\documentclass[
reprint,
superscriptaddress,
 amsmath,amssymb,
 aps,
]{revtex4-2}

\usepackage[caption=false]{subfig} 
\usepackage{graphicx}
\usepackage{dcolumn}
\usepackage{bm}
\usepackage{mathtools} 
\usepackage{hyperref}
\usepackage{graphicx,color,hyperref,xcolor}

\usepackage{braket} 

\newcommand{\ETHZ}{Institute for Theoretical Studies, ETH Z\"urich, Scheuchzerstrasse 70, 8006 Z\"urich, Switzerland}

\begin{document}


\title{Color Centers and Hyperbolic Phonon Polaritons in Hexagonal Boron Nitride:\\ A New Platform for Quantum Optics}

\author{Jie-Cheng Feng}
\affiliation{Department of Physics, ETH Zürich, 8093 Z\"urich, Switzerland}
\affiliation{Max Planck Institute for the Structure and Dynamics of Matter, Center for Free Electron Laser Science, 22761 Hamburg, Germany}
\affiliation{Institut für Theorie der Statistischen Physik, RWTH Aachen University, 52056 Aachen, Germany}

\author{Johannes Eberle}
\affiliation{Institute of Quantum Electronics, ETH Z\"urich, 8093 Z\"urich, Switzerland}

\author{Sambuddha Chattopadhyay}
\affiliation{Institute for Theoretical Physics, ETH Z\"urich, 8093 Z\"urich, Switzerland}
\affiliation{Lyman Laboratory, Department of Physics, Harvard University, Cambridge, Massachusetts 02138, USA}

\author{Johannes Knörzer}
    \affiliation{Department of Physics, ETH Zürich, 8093 Z\"urich, Switzerland}
 	\affiliation{\ETHZ}

\author{Eugene Demler}
\affiliation{Institute for Theoretical Physics, ETH Z\"urich, 8093 Z\"urich, Switzerland}

\author{Ata\c{c} \.{I}mamo\u{g}lu}
\affiliation{Institute of Quantum Electronics, ETH Z\"urich, 8093 Z\"urich, Switzerland}

\date{\today}

\begin{abstract}
Hyperbolic phonon polaritons (HPPs) in hexagonal boron nitride (hBN) confine mid-infrared light to deep-subwavelength scales and may offer a powerful route to strong light-matter interactions.
Generation and control of HPPs are typically accessed using classical near-field probes, which limits experiments at the quantum level.
A complementary frontier in hBN research focuses on color centers: bright, stable, atomically localized emitters that have rapidly emerged as a promising platform for solid-state quantum optics.
Here we establish a key connection between these two directions by developing a cavity-QED framework in which a single hBN color center serves as a quantum source of HPPs. 
We quantify the emitter-HPP interaction and analyze two generation schemes.
The first is spontaneous emission into the phonon sideband, which can produce single-HPP events and, in ultrathin slabs, becomes single-mode with an enhanced decay rate.
The second is a stimulated Raman process that provides frequency selectivity, tunable conversion rate, and narrowband excitation. 
This drive launches spatially confined, ray-like HPPs that propagate over micrometer distances.
We also outline a two-emitter correlation measurement that can directly test the single-polariton character of these emissions.
By connecting color-center quantum optics with hyperbolic polaritonics, our approach enables quantum emitters to act as on-chip quantum sources and controls for HPPs, while HPPs provide long-range channels that couple spatially separated emitters.
Together, these capabilities point to a new direction for mid-infrared photonic experiments that unite strong coupling, spectral selectivity, and spatial reach within a single material system.
\end{abstract}

\maketitle

\section{Introduction}

Hyperbolic media are anisotropic materials whose principal dielectric components take opposite signs, yielding open, hyperbolic isofrequency contours.
This unconventional dielectric response supports electromagnetic modes with very large wavevectors, deep subwavelength confinement, and highly directional energy flow \cite{poddubnyHyperbolicMetamaterials2013,guoHyperbolicMetamaterialsDispersion2020,leeHyperbolicMetamaterialsFusing2022,podolskiyStronglyAnisotropicWaveguide2005,jacobOpticalHyperlensFarfield2006,galfskyActiveHyperbolicMetamaterials2015}.
In van der Waals crystals and heterostructures, hyperbolic dispersion can arise naturally from the coupling of light to anisotropic collective excitations \cite{basovPolaritonsVanWaals2016,lowPolaritonsLayeredTwodimensional2017,basovPolaritonPanorama2020,basovPolaritonicQuantumMatter2025}.
A notable example is hexagonal boron nitride (hBN): in its mid-infrared (mid-IR) Reststrahlen bands, hybridization of photons with anisotropic optical phonons gives rise to hyperbolic phonon polaritons (HPPs) \cite{caldwellSubdiffractionalVolumeconfinedPolaritons2014,caldwellPhotonicsHexagonalBoron2019,huPhononPolaritonsHyperbolic2020,suFundamentalsEmergingOptical2024}.

Recently, HPPs have drawn considerable attention as a novel platform for studying light–matter interactions \cite{ashidaCavityQuantumElectrodynamics2023,narimanovHyperbolicQuantumProcessor2024,andolinaQuantumElectrodynamicsGraphene2025a,kerenCavityalteredSuperconductivity2026}.
The deep subwavelength confinement with large density of states substantially enhances the light-matter coupling, offering a practical route to the strong and even ultrastrong coupling regimes.
Compared with many plasmonic implementations, HPPs in high-quality hBN also exhibit intrinsically lower losses in the mid-IR \cite{khurginHowDealLoss2015,niLongLivedPhononPolaritons2021,gilesUltralowlossPolaritonsIsotopically2018,herzigsheinfuxHighqualityNanocavitiesMultimodal2024}.
These advantages make HPPs a promising setting for research in cavity quantum materials, in which the engineered photonic vacuum modifies the properties of the material \cite{sentefCavityQuantumelectrodynamicalPolaritonically2018,garcia-vidalManipulatingMatterStrong2021,schlawinCavityQuantumMaterials2022}.

Nevertheless, excitation and control of HPPs remain experimentally challenging.
As the relevant momenta far exceed those of free-space photons, most studies employ near-field techniques, particularly scattering-type scanning near-field optical microscopy (s-SNOM), where a metal tip converts incident radiation into large-$k$ polaritons and images the resulting interference patterns \cite{daiTunablePhononPolaritons2014a,shiAmplitudePhaseResolvedNanospectral2015,yoxallDirectObservationUltraslow2015,liHyperbolicPhononpolaritonsBoron2015,daiSubdiffractionalFocusingGuiding2015a}.
While SNOM has become the standard technique for launching and probing HPPs, it remains inherently classical and relies on an external probe, limiting the exploration of intrinsic quantum light-matter interactions.

In recent years, a wide variety of optically active defects in hBN have been discovered, spanning the full visible spectrum \cite{fournierPositioncontrolledQuantumEmitters2021, galeSiteSpecificFabricationBlue2022, fartasReproducibleGenerationGreenemitting2025, kumarLocalizedCreationYellow2023, whitefield2025generation}. Recent studies have demonstrated the reproducible creation of these defects in few-nanometer thick hBN \cite{liuSinglePhotonEmitters2025} down to monolayers \cite{tranQuantumEmissionHexagonal2016a}. Their high brightness, exceptional photostability and demonstrated photon indistinguishability \cite{gerardResonanceFluorescenceIndistinguishable2025}  make them promising candidates for next-generation quantum light sources, nanoscale sensors, and solid-state information processing platforms.

\begin{figure}
\centering
\includegraphics[width=0.8\linewidth]{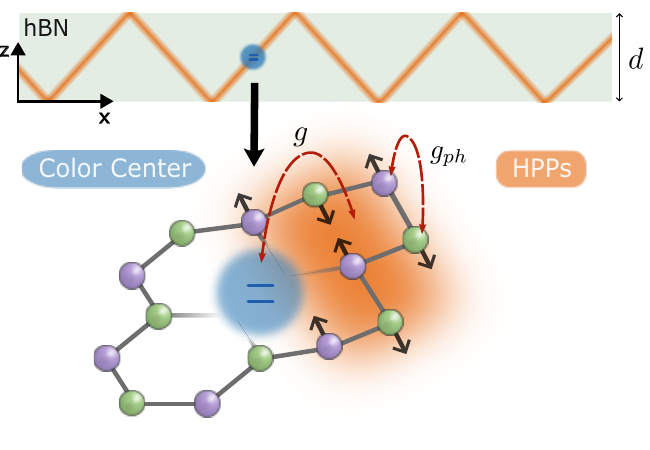}
\caption{\label{fig:setup}
Schematic of the hyperbolic phonon polariton (HPP) cavity considered in this work.
Photons and optical phonons in an hBN slab (light green) hybridize with coupling strength $g_{ph}$, forming HPPs (orange rays) that propagate inside the slab and decay evanescently outside.
HPPs can also interact with a localized color center (blue sphere) embedded in the hBN, with coupling strength $g$. 
The color center is modeled as a two-level system. 
The slab, of thickness $d$, is assumed to be surrounded by air.
The lower panel shows a zoomed-in microscopic illustration of the local lattice environment around the color center.
Phonons are represented by lattice vibrations, while photon-mediated interactions are indicated by the red dashed curves.
}
\end{figure}

The subnanometer spatial extent of color centers enables coupling to high-momentum electromagnetic fields, and several centers exhibit a pronounced mid-IR response in their emission spectra, indicating a strong interaction with lattice vibrations \cite{shevitskiBluelightemittingColorCenters2019,fournierPositioncontrolledQuantumEmitters2021,galeSiteSpecificFabricationBlue2022,fournierInvestigatingFastSpectral2023,liuSinglePhotonEmitters2025}.
Motivated by these, we establish a cavity quantum electrodynamics (cavity QED) framework that couples localized color centers (quantum emitters) to confined HPP modes (Fig.~\ref{fig:setup}) and show how a single emitter can function as a compact quantum source of HPPs.
We analyze two complementary generation mechanisms.
Spontaneous emission into the phonon sideband (PSB) can produce single-HPP events, whose rates and features depend on the hBN slab thickness.
A stimulated Raman scheme adds spectral and dynamical control: the second laser fixes the polariton frequency, sets the transition rate through its amplitude, and, by narrowing the linewidth, produces HPPs that propagate directionally as ray-like fields over micrometer distances.

Connecting these two rapidly developing lines of research, color centers and HPPs, is valuable for two reasons.
First, the quantum nature of the emitter provides access to nonclassical states of the polaritonic field that go beyond what can be studied using near-field techniques.
Second, the long-range propagation of HPPs offers a channel to mediate interactions between spatially separated emitters, enabling controlled exchange of energy and, in the quantum regime, transfer of states and entanglement over micron-scale distances.
Together, these capabilities suggest a new direction for mid-IR light–matter experiments that combine strong coupling, spectral selectivity, and spatial reach within a single material system.

The paper is organized as follows. 
In Sec.~[\ref{sec:HPP}], we introduce HPPs in thin hBN slabs and quantize their electromagnetic field, highlighting the vacuum field strength as a function of frequency and momentum.  
In Sec.~[\ref{sec:coupling}], we develop the interaction Hamiltonian between HPPs and localized color centers, and propose two mechanisms for HPP generation: spontaneous emission via the PSB, for which we also show consistent experimental results, and stimulated Raman excitation. 
Sec.~[\ref{sec:propagation}] analyzes the propagation of the generated HPPs, showing that narrowband excitation produces directional, ray-like fields over micrometers.  
In Sec.~[\ref{sec:conclusion}], we present a summary of results and suggest several interesting directions for future research.

\section{Hyperbolic Phonon Polaritons in an hBN Cavity}
\label{sec:HPP}

In this section, we first provide a brief overview of HPPs in hBN. 
We then introduce the geometry of our system, a thin hBN slab supporting confined HPP modes.
We also discuss the quantization of these modes and compute the vacuum electric field fluctuations that underlie their interaction with localized emitters.

\subsection{Hyperbolic Materials}

\begin{figure}[!b]
\centering
\includegraphics[width=0.9\linewidth]{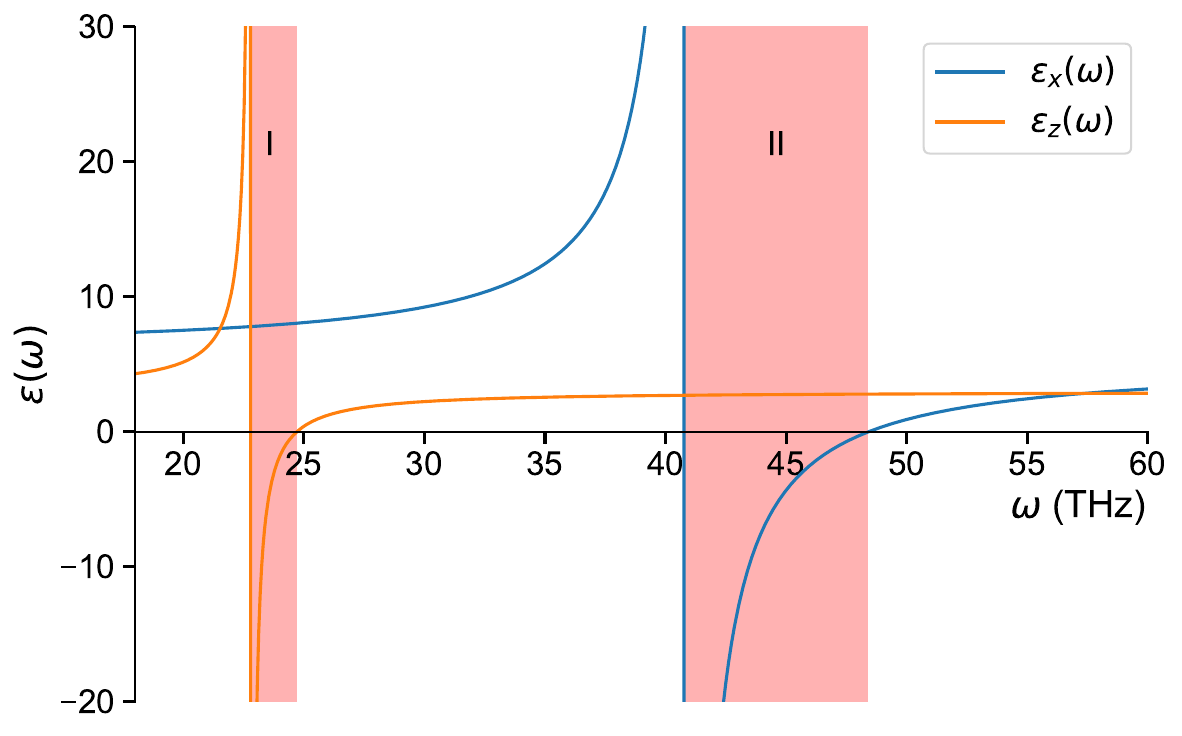}
\caption{\label{fig:dielectric plot}Dielectric function of hBN.
The blue and orange lines denote the in-plane and out-of-plane components, respectively.
The red shaded areas indicate the two Reststrahlen bands, where one component of the dielectric function is negative and the material becomes hyperbolic. }
\end{figure}

Hyperbolic materials exhibit a different electromagnetic response compared to conventional dielectrics \cite{poddubnyHyperbolicMetamaterials2013,guoHyperbolicMetamaterialsDispersion2020,leeHyperbolicMetamaterialsFusing2022,smithElectromagneticWavePropagation2003,podolskiyStronglyAnisotropicWaveguide2005,veselagoLeftHandBrightness2006,jacobEngineeringPhotonicDensity2010,jacobBroadbandPurcellEffect2012}.
From Maxwell’s equations in a source-free, uniaxial, nonmagnetic medium, we can solve the dispersion relation of the transverse-magnetic (TM) polarized mode, where the magnetic field $\boldsymbol{B}$ is purely in-plane ($\boldsymbol{B}\cdot\hat{\boldsymbol{z}}=0$),
\begin{eqnarray}
\frac{k_x^2 + k_y^2}{\epsilon_z(\omega)} + \frac{k_z^2}{\epsilon_x(\omega)} = \frac{\omega^2}{c^2},
\label{eq:TM_dispersion}
\end{eqnarray}
where $\epsilon_x$($\epsilon_z$) is the in-plane (out-of-plane) diagonal component of the dielectric tensor, typically frequency-dependent.
Without loss of generality, we set $k_y=0$ in the following discussion.
When $\epsilon_x \epsilon_z <0$, this TM mode becomes hyperbolic, and the real solution exists even in the deep subwavelength limit $k_x, k_z \gg \omega/c$, since the two terms can cancel each other on the left-hand side of Eq.~\eqref{eq:TM_dispersion}.
This property allows hyperbolic materials to support multiple highly confined, propagating modes beyond the diffraction limit, making them a unique platform for subwavelength optics.
In fact, the ratio between $k_x$ and $k_z$, named $\kappa$ in this work, is fixed in the subwavelength limit,
\begin{eqnarray}
\kappa(\omega) := \frac{k_x}{k_z} \approx \sqrt{-\frac{\epsilon_z(\omega)}{\epsilon_x(\omega)}}.
\label{eq:kxkz_ratio}
\end{eqnarray}

Arguably, hBN is the most extensively studied natural hyperbolic material \cite{caldwellPhotonicsHexagonalBoron2019,huPhononPolaritonsHyperbolic2020,suFundamentalsEmergingOptical2024}.
Its hyperbolic dielectric response originates from the anisotropic phonon frequencies, and these hybridized light–matter excitations are referred to as HPPs \cite{daiTunablePhononPolaritons2014a,caldwellSubdiffractionalVolumeconfinedPolaritons2014,shiAmplitudePhaseResolvedNanospectral2015,yoxallDirectObservationUltraslow2015,lowPolaritonsLayeredTwodimensional2017,kurmanSpatiotemporalImaging2D2021}.

In polar materials, the frequency range between the longitudinal optical (LO) and transverse optical (TO) phonon is generally referred to as the Reststrahlen band, within which the dielectric function becomes negative.
Due to the 2D nature of van der Waals materials like hBN, the relevant physical parameters differ between the in-plane and out-of-plane directions, leading to two separate Reststrahlen bands, in which only one component of the dielectric function becomes negative \cite{geickNormalModesHexagonal1966,kernInitiocalculationLatticeDynamics1999,reichResonantRamanScattering2005,serranoVibrationalPropertiesHexagonal2007,michelPhononDispersionsPiezoelectricity2011}.
The dielectric function of hBN along the $i\in\{x,z\}$ direction is
\begin{eqnarray}
    \epsilon_i(\omega)=\epsilon_{i\infty}\frac{\Omega^2_{i,\rm LO}-\omega^2-i\gamma_{i}\omega}{\Omega^2_{i,\rm TO}-\omega^2-i\gamma_{i}\omega},
    \label{eq:dielectric_function}
\end{eqnarray}
which is shown in Fig.~\ref{fig:dielectric plot} with $\Omega_{x,\rm TO} = 1360\ \text{cm}^{-1}$, $\Omega_{x,\rm LO} = 1614\ \text{cm}^{-1}$, $\Omega_{z,\rm TO} = 760\ \text{cm}^{-1}$, $\Omega_{z,\rm LO} = 825\ \text{cm}^{-1}$, and $\epsilon_{x\infty} = 4.9$, $\epsilon_{z\infty} = 2.95$ \cite{suFundamentalsEmergingOptical2024}.
In this work, we neglect the damping term $\gamma_i$ in the Lorentzian model for calculating the light-matter interaction, as the optical losses in hBN are relatively small \cite{khurginHowDealLoss2015,gilesUltralowlossPolaritonsIsotopically2018,niLongLivedPhononPolaritons2021}.

\subsection{Thin slab of hBN as a two dimensional cavity}

The setup we consider is a thin slab of hBN with a thickness $d$, as depicted in Fig.~\ref{fig:setup}, where the wavevector is real inside but the field is evanescent outside. 
Therefore, this geometry effectively turns the slab into an optical cavity \cite{ashidaCavityQuantumElectrodynamics2023}.
The eigenmodes can be found by solving Maxwell’s equations with appropriate boundary conditions (Appendix \ref{app:eigenmodes}).
In the deep subwavelength limit, the HPP modes satisfy:
\begin{eqnarray}
k_x = \frac{\kappa}{d} \Bigr[ 2 \text{tan}^{-1}(-\frac{1}{\epsilon_x\kappa}) + n\pi \Bigr],
\label{eq:HPP_resonance}
\end{eqnarray}
where $n$ is the quantum number arising from the confinement along the
z-direction and $\kappa$ is from Eq.~\eqref{eq:kxkz_ratio}.
Since $\epsilon_x(\omega)$ and $\epsilon_z(\omega)$ are frequency dependent, this condition dictates the dispersion of HPPs in the slab.
In this work, we focus on the second Reststrahlen band between 40 to 50 THz.
Notice that this dispersion is valid for an intermediate momentum regime, where $k$ is much larger than the free-space photon momentum $\omega/c$ but still much smaller than reciprocal lattice vectors, so that phonon dispersion can be neglected.
Extending the description to higher momenta would require incorporating momentum dependence into the dielectric function or adopting a fully microscopic treatment.

\begin{figure}[t]
  \centering

  \subfloat[\label{fig:dispersion_HPP}]{%
    \includegraphics[width=0.9\linewidth]{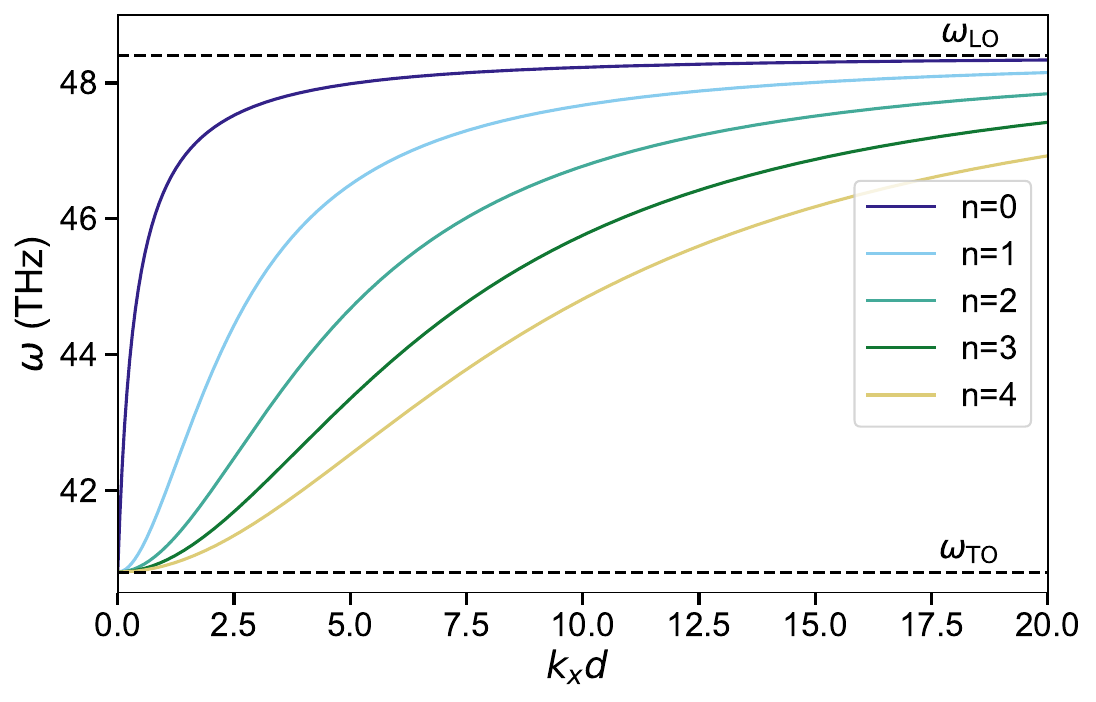}%
  }

  \vspace{-0.5cm} 

  \subfloat[\label{fig:vacuum_strength}]{%
    \includegraphics[width=0.9\linewidth]{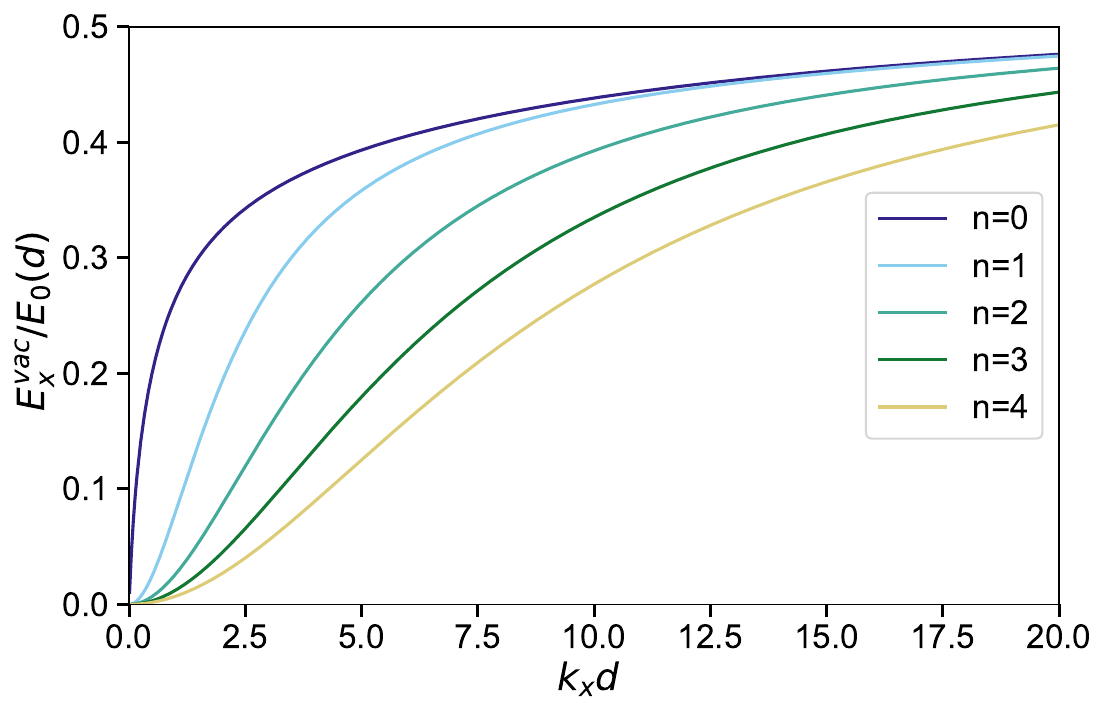}%
  }

  \caption{\label{fig:HPP_combined}
The dispersion and vacuum field strength of HPPs (of the second Reststrahlen band) in an hBN slab.
From top to bottom, the solid curves correspond to the mode indices $n=0,1,2,3,4$, respectively.
(a) Dispersion of HPPs in an hBN slab surrounded by air.
The black dashed lines indicate the in-plane LO and TO phonon frequencies.
(b) The in-plane vacuum electric field strength of HPPs in an hBN slab, normalized by $E_0(d) = \sqrt{\frac{\hbar\omega}{2\epsilon_0\epsilon_{x\infty}d A_{\text{eff}}}}$.}
\end{figure}

The resulting dispersion of HPP modes is shown in Fig.~\ref{fig:dispersion_HPP}. 
When the frequency approaches \( \omega_{TO} \), the ratio \( \kappa \to 0 \), corresponding to a very small in-plane momentum \( k_x \). 
In contrast, as the frequency approaches \( \omega_{LO} \), \( \kappa \) and \( k_x \) become large. 
Higher-order modes (\( n > 0 \)) at fixed frequency correspond to increasingly larger momenta.

Next, we need to quantize the HPPs properly and compute the vacuum fluctuation of their electric field inside the hBN cavity.
The standard approach in cavity QED to quantize the electromagnetic field involves normalizing the energy of classical eigenmodes to that of a single photon at the mode frequency, $\hbar\omega$.
However, applying it to electromagnetic fields in materials is not always straightforward due to the complex interactions between the field and the matter’s degrees of freedom \cite{glauberQuantumOpticsDielectric1991,santosElectromagneticfieldQuantizationInhomogeneous1995,feistMacroscopicQEDQuantum2020a}.
To address this, theoretical frameworks such as macroscopic QED have been developed \cite{bergmanSurfacePlasmonAmplification2003,knollQEDDispersingAbsorbing2003,scheelMacroscopicQEDConcepts2009a,philbinCanonicalQuantizationMacroscopic2010a}.

With either the macroscopic QED or a microscopic Hamiltonian of HPPs (Appendix \ref{app:quantization}), we can quantize the vector potential in the following form:
\begin{eqnarray}
\boldsymbol{\hat{A}}(\boldsymbol{r})=\sum_m \sqrt{\frac{\hbar}{2\epsilon_0\omega_m}}[\boldsymbol{F}_m(\boldsymbol{r})\boldsymbol{\hat{\gamma}}_m +\boldsymbol{F}^*_m(\boldsymbol{r})\boldsymbol{\hat{\gamma}}^\dagger_m],
\label{eq:quantized_A}
\end{eqnarray}
where $\boldsymbol{F}_m$ denotes the $m$-th mode function of vector potential with mode frequency $\omega_m$, and $\boldsymbol{\hat{\gamma}}^\dagger_m(\boldsymbol{\hat{\gamma}}_m)$ is the corresponding polariton creation (annihilation) operator.

Figure~[\ref{fig:vacuum_strength}] displays the in-plane vacuum electric field strength \( E^{\mathrm{\rm vac}}_x \) of HPP modes in an hBN slab as a function of the in-plane momentum, for different mode indices \( n \). 
The field amplitude is normalized to \( E_0(d) = \sqrt{\hbar\omega / (2\epsilon_0 \epsilon_{x\infty} d A_{\text{eff}})} \), where $A_{\text{eff}}$ is the effective in-plane area.
In the long-wavelength limit \( k_x \to 0 \), the electromagnetic mode becomes delocalized, with the field extending far beyond the slab. 
As a result, the vacuum fluctuation of the electric field is weak.
As the in-plane momentum increases, the field becomes increasingly confined within the hBN slab due to exponential decay outside the slab boundaries.
This confinement leads to a significant enhancement of the vacuum electric field amplitude.


\section{Hyperbolic Phonon Polariton Generation via Color Centers}
\label{sec:coupling}

\begin{figure}[!b]
\centering

\subfloat[\label{fig:illus_spontaneous}]{%
    \includegraphics[width=0.45\linewidth]{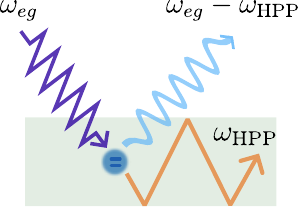}%
}%
\hspace{0.04\linewidth}%
\subfloat[\label{fig:illus_stimulated}]{%
    \includegraphics[width=0.45\linewidth]{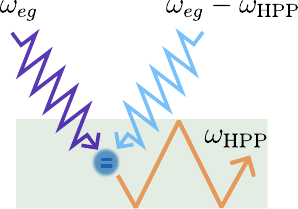}%
}%

\caption{\label{fig:illus_generation_combined}
 Schematic of the two mechanisms for generating HPPs from a single color center in an hBN slab. 
(a) Spontaneous phonon sideband emission: an optical excitation at frequency $\omega_{eg}$ (purple) relaxes by emitting one HPP at $\omega_{\rm HPP}$ (orange) together with a Stokes photon at $\omega_{eg}-\omega_{\rm HPP}$ (blue). 
(b) Stimulated Raman process: two drives at $\omega_{eg}$ and $\omega_{eg}-\omega_{\rm HPP}$ stimulates the transition, leading to narrowband, coherent HPP emission at $\omega_{\rm HPP}$.
}
\end{figure}

In this section, we examine the interaction between HPPs and localized emitters embedded in hBN. 
Particularly promising classes of such emitters are lattice defects, commonly referred to as color centers \cite{suFundamentalsEmergingOptical2024,kimColorCentersHexagonal2023,cholsukHBNDefectsDatabase2024}. 
These defects can be modeled as two-level systems with excitation energies on the order of several eV, as indicated by their zero-phonon lines (ZPLs) in photoluminescence (PL) measurements.
Because these energies are much higher than the characteristic mid-infrared frequencies of HPPs, direct resonant coupling of the Jaynes–Cummings type is not expected.
However, many color centers exhibit strong phonon sidebands (PSBs) in their PL spectra, such as the  Blue Center (B-center) \cite{galeSiteSpecificFabricationBlue2022,fournierPositioncontrolledQuantumEmitters2021,liuSinglePhotonEmitters2025}.
These PSBs occur at frequencies overlapping with the Reststrahlen band, which suggests the possibility of generating HPPs during the de-excitation of the emitter via phonon-assisted processes \cite{vuongPhononPhotonMappingColor2016,wiggerPhononassistedEmissionAbsorption2019,khatriPhononSidebandsColor2019}.
In the following, we propose two mechanisms for generating HPPs using the color centers shown in Fig.~\ref{fig:illus_generation_combined} : one via this spontaneous emission in PSBs in Sec.~\ref{ssec:hpp-via-spontaneous-emission}, and the other via the resonant stimulated Raman scattering in Sec.~\ref{ssec:hpp-via-raman}.

\subsection{HPP Generation via Spontaneous Phonon Sidebands Emission}
\label{ssec:hpp-via-spontaneous-emission}

To interpret the PSB, we model the emitter as a dipole with moment $\boldsymbol{\mu}_d$, coupled to the electric field of the HPPs, analogous to Fröhlich-type electron-phonon coupling \cite{frohlichElectronsLatticeFields1954,yuFundamentalsSemiconductorsPhysics2010,wiggerPhononassistedEmissionAbsorption2019}. 
The total Hamiltonian for an emitter located at position $\boldsymbol{r}=0$ is given as 

\begin{eqnarray}
\hat{H}_{\rm vac}=\hat{H}_0+\hat{H}_{\rm int},\nonumber
\end{eqnarray}
with a bare Hamiltonian
\begin{flalign}
&\hat{H}_0/\hbar = \omega_{eg} \hat{\sigma}_{ee} +\sum_{\boldsymbol{q}}\omega_{\boldsymbol{q}}\hat{a}^\dagger_{\boldsymbol{q}}\hat{a}_{\boldsymbol{q}}+ \sum_{\boldsymbol{k_x},n}\omega_{\boldsymbol{k_x},n}\hat{\gamma}^\dagger_{\boldsymbol{k_x},n}\hat{\gamma}_{\boldsymbol{k_x},n}\nonumber ,
\end{flalign}
and an interaction part
\begin{flalign}
&\hat{H}_{\rm int}/\hbar=\sum_{\boldsymbol{q}} g_{\boldsymbol{q}}(\hat{\sigma}_{eg}+\hat{\sigma}_{ge})(\hat{a}_{\boldsymbol{q}} + \hat{a}^\dagger_{\boldsymbol{q}}) \nonumber\\
&+\sum_{\boldsymbol{k_x},n} \frac{\boldsymbol{\mu}_d } {\hbar}  \hat{\sigma}_{ee} \cdot \sqrt{\frac{\hbar\omega_{\boldsymbol{k_x},n}}{2\epsilon_0}}[\boldsymbol{F}_{\boldsymbol{k_x},n}(0)\hat{\gamma}_{\boldsymbol{k_x},n} +h.c.],
\end{flalign}
where $\omega_{eg}$ is the transition frequency between the excited and ground states of the emitter, and
$\hat{\sigma}_{ij}=|i\rangle\langle j|$ $(i,j\in\{g,e\})$ are the emitter operators, such that
$\hat{\sigma}_{eg}$ ($\hat{\sigma}_{ge}$) is the raising (lowering) operator and $\hat{\sigma}_{ee}$ is the excited-state population operator.
$\omega_q,\hat{a}_q,\hat{a}_q^\dagger$ denote the frequencies and annihilation/creation operators of the optical photon,
while $\omega_{k_x,n},\hat{\gamma}_{\boldsymbol{k_x},n},\hat{\gamma}_{\boldsymbol{k_x,}n}^\dagger$ are the corresponding quantities for the THz HPPs.
Here $g_{\boldsymbol{q}}$ denotes the optical photon-emitter coupling strength, and the coupling strength between HPPs and emitter ($g$ in Fig.~\ref{fig:setup}) is $\sqrt{\frac{\hbar\omega_{\boldsymbol{k_x},n}}{2\epsilon_0}}\boldsymbol{\mu}\cdot\boldsymbol{F}_{\boldsymbol{k_x},n}=\boldsymbol{\mu}\cdot\boldsymbol{E}^{\rm vac}$.
Assuming the coupling is dominantly in-plane \cite{maciaszekBlueQuantumEmitter2024}, the coupling strength between emitter and HPPs in thin hBN slab is $\mu_d E^{\rm vac}_x$, where $E^{\rm vac}_x$ is presented in Fig.~\ref{fig:vacuum_strength}.

Since the vacuum fluctuation is typically weak, we can perform a Schrieffer–Wolff (SW) transformation to find the effective Hamiltonian $H'$.
We choose the generator
\begin{eqnarray}
S=-\sum_{\boldsymbol{k_x},n} \frac{\mu_d E^{\rm vac}_x}{\hbar\omega_{\boldsymbol{k_x},n}}\hat{\sigma}_{ee} [ \hat{\gamma}_{\boldsymbol{k_x},n} - \hat{\gamma}^{\dagger}_{\boldsymbol{k_x},n}].
\label{eq:SW1}
\end{eqnarray}
The effective Hamiltonian is then approximated as
\begin{eqnarray}
&&\hat{H}'_{\rm vac} = e^S \hat{H}_{\rm vac} e^{-S} \nonumber\\
&&\approx \hat{H}_0 - \sum_{\boldsymbol{k_x},n} \frac{(\mu_d  E^{\rm vac}_x)^2}{\hbar\omega_{\boldsymbol{k_x},n}}\hat{\sigma}_{ee} \nonumber \\
&&+\sum_{\boldsymbol{q}} g_{\boldsymbol{q}}(\hat{\sigma}_{eg}+\hat{\sigma}_{ge})(\hat{a}_{\boldsymbol{q}} + \hat{a}^\dagger_{\boldsymbol{q}})\nonumber\\&&
-\sum_{\boldsymbol{k_x},n,\boldsymbol{q}} \frac{\mu_d E^{\rm vac}_x g_{\boldsymbol{q}}}{\hbar\omega_{\boldsymbol{k_x},n}}(\hat{\sigma}_{eg}-\hat{\sigma}_{ge})(\hat{a}_{\boldsymbol{q}} + \hat{a}^\dagger_{\boldsymbol{q}}) (\hat{\gamma}_{\boldsymbol{k_x},n}-\hat{\gamma}^{\dagger}_{\boldsymbol{k_x},n})\nonumber\\
\label{eq:SW2}
\end{eqnarray}

Applying the rotating wave approximation (RWA) to the effective model in Eq.~\eqref{eq:SW2}, the ZPL corresponds to the term $\hat{\sigma}_{ge}\hat{a}^\dagger_{\boldsymbol{q}}$ with frequency $\omega_{\boldsymbol{q}} \approx \omega_{eg}$, and the PSB, illustrated in Fig.~\ref{fig:illus_spontaneous}, corresponds to the term $\hat{\sigma}_{ge}\hat{a}^\dagger_{\boldsymbol{q}}\hat{\gamma}^\dagger_{\boldsymbol{k_x},n}$ with frequency $\omega_{\boldsymbol{q}}\approx\omega_{eg}-\omega_{\boldsymbol{k_x},n}$.

This confirms that the coupling between the emitter and HPPs can lead to a PSB within the frequency range of the Reststrahlen band of hBN (approximately 160–200 meV).
To quantify the strength of this coupling, we consider the intensity ratio between PSB and ZPL:
\begin{eqnarray}
\frac{A_{\rm PSB}}{A_{\rm ZPL}}&&=\frac{\Gamma_{\rm PSB}}{\Gamma_{\rm ZPL}}
\approx \sum_{\boldsymbol{k_x},n,\boldsymbol{q}}\Bigr|\frac{\mu_d E^{\rm vac}_x g_{\boldsymbol{q}}} {\hbar\omega_{\boldsymbol{k_x},n}}\Bigr|^2 \Bigr/ \Bigr| \sum_{\boldsymbol{q}} g_{\boldsymbol{q}}\Bigr|^2 \nonumber\\
&&\approx \sum_{\boldsymbol{k_x},n}\Bigr|\frac{\mu_d E^{\rm vac}_x} {\hbar\omega_{\boldsymbol{k_x},n}}\Bigr|^2,
\label{eq:PSB_ZPL_ratio}
\end{eqnarray}
where the transition rates $\Gamma$ are computed using Fermi’s golden rule.
Since $\hbar g=\mu_d E^{\rm vac}_x$ in our setup, this ratio is proportional to $(g/\omega_{\rm HPP})^2$.
Thus, it can serve as a measure of the strength of the light-matter interaction and an indicator of the (ultra)strong coupling regime \cite{friskkockumUltrastrongCouplingLight2019,forn-diazUltrastrongCouplingRegimes2019,garcia-vidalManipulatingMatterStrong2021}.

Notice that, in hyperbolic materials, this sum can diverge without a cutoff $\Lambda$ of wavevector $k_x$.
In our case, a natural physical cutoff is given by the size of the emitter, $l_e$, leading to $\Lambda\approx1/l_e$ \cite{poddubnySpontaneousRadiationFinitesize2011}.
The cutoff of the HPP branch can be estimated in a similar way $N_0 \equiv n_{\text{max}}\approx\lfloor \frac{\Lambda d}{\pi} \rfloor$.

\begin{figure}
\centering
\includegraphics[width=0.9\linewidth]{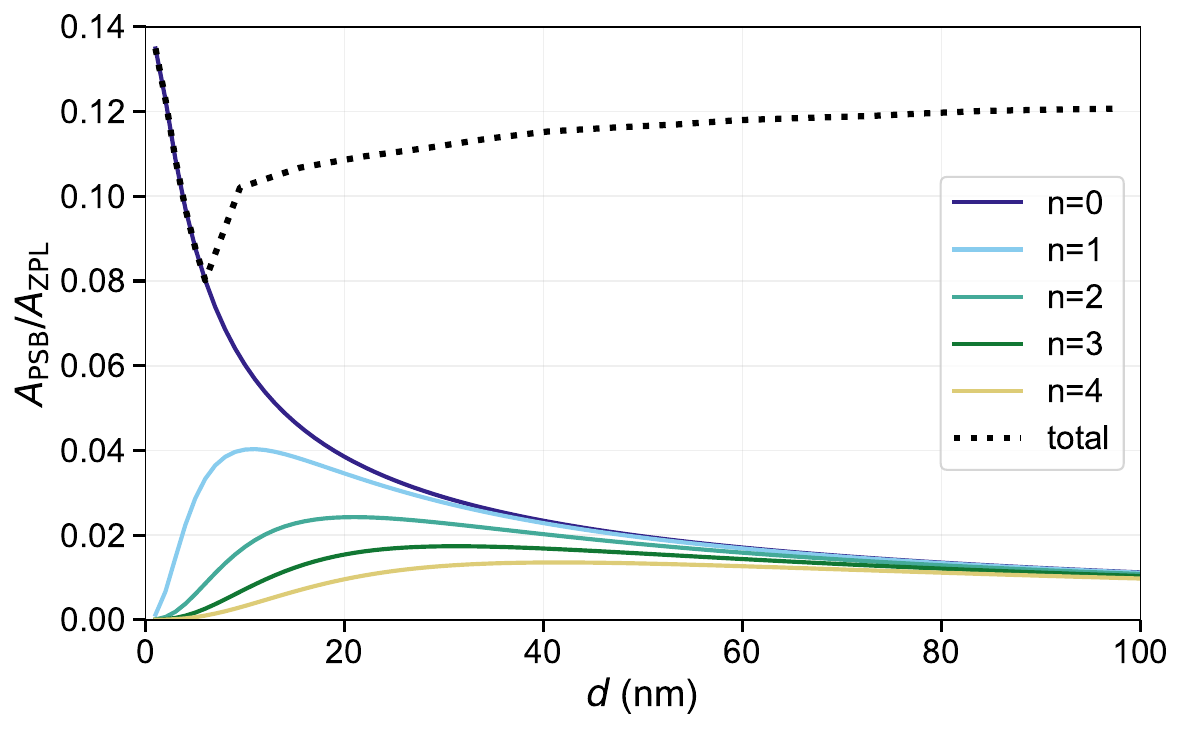}
\caption{\label{fig:PSB} 
Calculated intensity ratio between the phonon sideband (PSB) and the zero-phonon line (ZPL) as a function of slab thickness. 
We use a momentum cutoff \( \Lambda = 1/(2\,\text{nm}) \) and dipole moment \( \mu_d = e_0 \cdot (2\,\text{nm}) \) for illustration. 
Solid curves correspond to HPP branches with mode indices \( n = 0, 1, 2, 3, 4 \) (from top to bottom).
The dotted curve sums the contributions of the branches $n=0$ up to $n\approx\lfloor \frac{\Lambda d}{\pi} \rfloor$, so fewer branches enter for thinner hBN.
As $d$ decreases, the coupling per branches first grows while higher-order branches gradually drop out due to the momentum cutoff.
In the ultrathin hBN, the response is dominated by the \(n=0\) branch, i.e., a single-mode HPP cavity.}
\end{figure}

In Fig.~\ref{fig:PSB}, we plot the PSB/ZPL intensity ratio using the cutoff $\Lambda=1/(2 \text{nm})$ and dipole moment $\mu_d=e_0(2\text{nm})$ for demonstration, where $e_0$ is the unit charge.
We show additional results with different parameter choices in Appendix~\ref{app:additional_PL}, and the overall trends remain the same.
The “total” curve sums the contributions from $n=0$ up to $n=N_0\approx\lfloor \frac{\Lambda d}{\pi} \rfloor$ branches, so that there are fewer branches included for thinner hBN.
Since $N_0$ varies discretely with $d$, some step-like changes remain even after averaging.

At large slab thicknesses, contributions from different branches converge and the total ratio becomes independent of $d$ (Eq.~\eqref{eq:LO_HPPS} in Appendix~\ref{app:quantization}).
Although the calculated PSB/ZPL ratio is somewhat smaller than what is often observed experimentally for B-centers, it is of the same order of magnitude ($10^{-1}$) and captures the correct scale of the effect.
The vacuum field per mode starts to increase and enhance PSB emission as the slab becomes thinner.
However, in very thin slabs, the couplings to $n>0$ HPP branches diminish because their in-plane momenta exceed the cutoff $\Lambda$.
Together with the decreasing number of branches included in the total PSB, this drives the system towards a regime where the $n=0$ HPP mode dominates.
Hence, in ultrathin hBN, a single-mode HPP response is expected and a clear PSB enhancement should be observable in PL, offering an experimental signature of the emitter-HPP interaction.

\begin{figure*}
    \centering
    \includegraphics[width=0.8\linewidth]{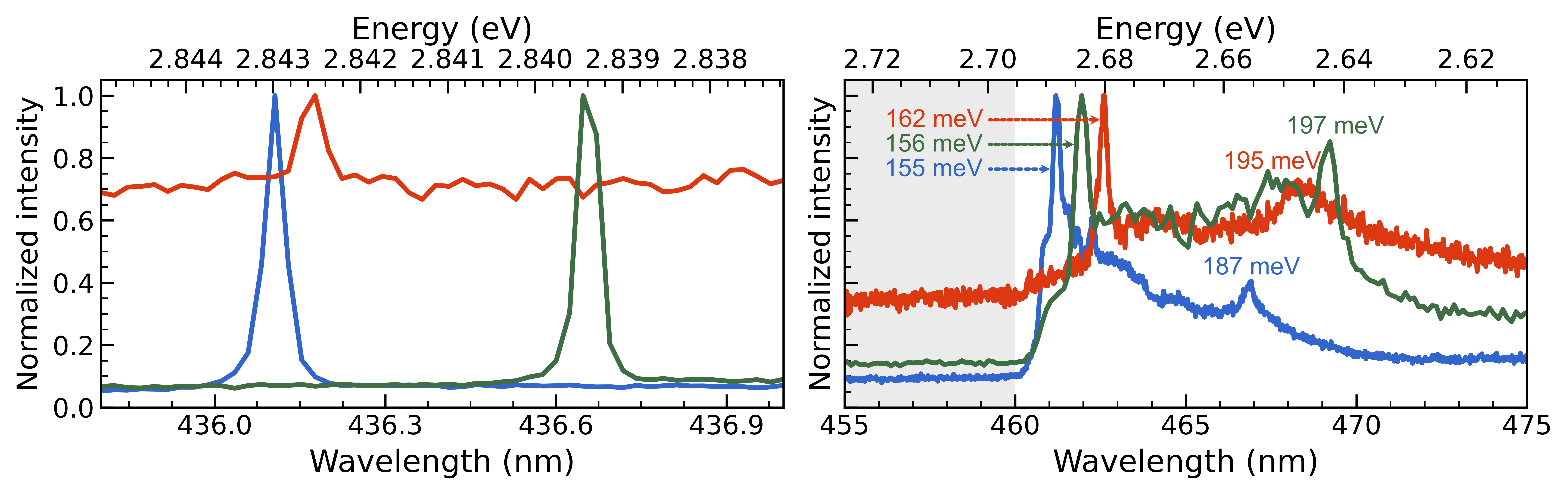}
    \caption{Experimental PL spectra of B-centers in three different hBN samples (colors).
    Left: ZPL measured under 406 nm excitation (normalized).
    Right: PSB measured under resonant excitation of the ZPL; the gray shaded region marks the blocked wavelength range of the long-pass filter used in the detection path. 
    Top axes indicate photon energy.
    For each sample, the energy separations between the ZPL and the two prominent PSB peaks are: blue, 155 and 187 meV; green, 156 and 197 meV; red, 162 and 195 meV.}
    \label{fig:PSB_spectra}
\end{figure*}

Beyond the overall strength, the spectral structure of the PSB can also be informative.
As shown in Figs.~\ref{fig:dispersion_HPP} and \ref{fig:vacuum_strength}, both the HPP density of states and the vacuum electric field increase as the frequency approaches the in-plane LO frequency, $\Omega_{\mathrm{LO}}\simeq 48.4~\mathrm{THz}$ ($\simeq 200~\mathrm{meV}$).
This suggests that polariton-assisted PSB emission should be strongest close to, but below, $200~\mathrm{meV}$ detuning.
To test this, we measured the PL of three different B-centers across three samples \cite{liuSinglePhotonEmitters2025}, and the data are shown in Fig.~\ref{fig:PSB_spectra} (experimental details in the Appendix~\ref{app:PL_setup}).
All three emitters show a sharp PSB peak at $155$--$162~\mathrm{meV}$ and a broader feature at $ 187$--$197~\mathrm{meV}$.
The PSB peak positions vary by a few meV across samples, which is likely due to the local strain of different samples.
The first peak near $160~\mathrm{meV}$ coincides with a maximum in the bare hBN phonon density of states, largely contributed by the optical phonons near the M and K points \cite{kernInitiocalculationLatticeDynamics1999,serranoVibrationalPropertiesHexagonal2007,vuongPhononPhotonMappingColor2016}.
In contrast, the bare phonon density of states is vanishing around the second peak, whereas our HPP density of states and vacuum field strength remain large, consistent with the interpretation of this peak as HPP-assisted PSB emission.
Additional discussion of the PSB peak positions is provided in Appendix~\ref{app:additional_PL}.



One of the key advantages of using color centers, compared to classical near-field probes, is their quantum nature.
As two-level systems, they not only provide the atomic-scale localization but can also serve as sources of nonclassical radiation, including single phonon polaritons.
While single-photon emission from hBN color centers has been demonstrated in the optical domain \cite{tranQuantumEmissionHexagonal2016a}, our framework suggests that spontaneous PSB emission can likewise produce a single HPP.
One simple test for single-polariton emission is to measure the second-order correlation function $g^{(2)}(\tau)$ of the PSB using a photon correlation setup.
However, additional information could be obtained using a two-emitter correlation measurement outlined below.

\begin{figure}[!b]
\centering
\includegraphics[width=0.8\linewidth]{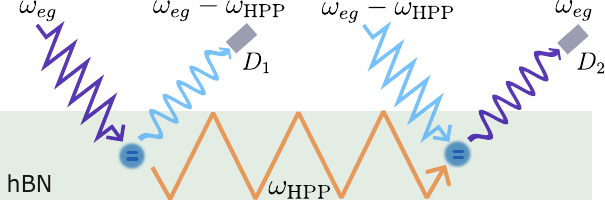}
\caption{\label{fig:two_center}Two color centers (blue spheres) are embedded in an hBN slab and separated by several micrometers.
The source emitter (left) is excited resonantly at $\omega_{eg}$(purple).
Its PSB emission at $\omega_{eg}-\omega_{\rm HPP}$ (blue) is recorded on detector $D_1$, while the same decay launches a HPP (orange) at frequency $\omega_{\rm HPP}$ that propagates inside the slab.
The detector emitter (right) is illuminated by a laser at $\omega_{eg}-\omega_{\text{HPP}}$, because its linewidth is far smaller than the THz polariton frequency scale, this drive can only excite the detector emitter if the arriving HPP provides the missing energy.
When excited, the detector emitter radiates into its ZPL at $\omega_{eg}$, collected on $D_2$.
Time-correlated counts between $D_1$ (PSB) and $D_2$ (ZPL) probe the quantum statistics of the launched polaritons; observation of antibunching in the correlation provides direct evidence of single-HPP emission.
}
\end{figure}

Consider two spatially separated color centers in hBN (their locations can be easily measured with an avalanche photodiode in experiments).
The first emitter is driven by a pulsed laser and emits into the PSB, which couples to the HPP modes.
The second emitter is illuminated with a narrowband Raman laser tuned to $\omega_{eg}-\omega_{HPP}$.
Because the Raman laser linewidth is typically much smaller than the THz HPP, excitation of the second emitter can only occur if an HPP generated by the first emitter reaches its position.
By recording time-resolved coincidences between detectors $D_1$ (monitoring the PSB of the source emitter) and $D_2$ (monitoring the ZPL of the second emitter), one can probe the quantum correlations of the emitted polaritons.
The experimental setup is illustrated in Fig.~\ref{fig:two_center}.

This setup effectively realizes a Hanbury Brown–Twiss–type measurement for phonon polaritons, where each color center serves as a local polariton detector.
Therefore, the color centers in hBN are not only the quantum \textit{sources} of HPPs, but also can be the quantum \textit{sensors}.
Demonstrating anti-bunching in this configuration would provide compelling evidence for single-HPP emission.
If single polaritons can be reliably generated and controlled within hBN, this would unlock a new solid-state platform for quantum optics, allowing direct implementation of single-photon–based protocols, such as interference, entanglement, and quantum information transfer, without relying on free-space photons.

Beyond probing antibunching, this time-resolved two-emitter measurement can also access the HPP group velocity via the delay between the source and detector signals.
This information could be relevant for future nonlinear quantum-optical applications, since a reduced group velocity increases the effective interaction time, thereby enhancing the optical nonlinearities at the level of single photons \cite{lukinNonlinearOpticsQuantum2000}.

\subsection{HPP Generation via Stimulated Raman Transition}
\label{ssec:hpp-via-raman}

Inspired by the possibility of generating HPPs from the PSB of the color centers, we now propose a stimulated version that produces HPPs in a steady, controllable, and narrowband fashion.
The idea, demonstrated in Fig.~\ref{fig:illus_stimulated}, is to illuminate a single color center with two monochromatic lasers: a pump at frequency $\omega_1=\omega_{eg}$ that addresses the $\ket{g}\leftrightarrow\ket{e}$ transition; a Raman laser at frequency $\omega_2 < \omega_{eg}$.
We will show that this process coherently converts the optical drive into a polariton of fixed frequency $\omega_{\rm HPP} = \omega_{eg} - \omega_2$ via stimulated Raman scattering.
The Hamiltonian for this scenario is:
\begin{eqnarray}
&&\hat{H}_{\rm sti}/\hbar= \omega_{eg}\hat{\sigma}_{eg} +\sum_{\boldsymbol{k_x},n}\omega_{\boldsymbol{k_x},n}\hat{\gamma}^\dagger_{\boldsymbol{k_x},n}\hat{\gamma}_{\boldsymbol{k_x},n} + \hat{H}_{\rm drive} , \nonumber\\
&& \hat{H}_{\rm drive} = [2\Omega_1\cos{(\omega_1 t)}+2\Omega_2\cos{(\omega_2 t)}](\hat{\sigma}_{eg}+\hat{\sigma}_{ge})\nonumber\\
&&+\sum_{\boldsymbol{k_x},n} \frac{\boldsymbol{\mu}_d } {\hbar}  \hat{\sigma}_{ee} \cdot \sqrt{\frac{\hbar\omega_{\boldsymbol{k_x},n}}{2\epsilon_0}}[\boldsymbol{F}_{\boldsymbol{k_x},n}(0)\hat{\gamma}_{\boldsymbol{k_x},n} +\boldsymbol{F}^*_{\boldsymbol{k_x},n}(0)\hat{\gamma}^\dagger_{\boldsymbol{k_x},n}],\nonumber\\
\label{eq:H_sti}
\end{eqnarray}
where the $\Omega_1$ ($\Omega_2$) represents the Rabi frequency of the first (second) laser.

\begin{figure}
\centering
\includegraphics[width=0.7\linewidth]{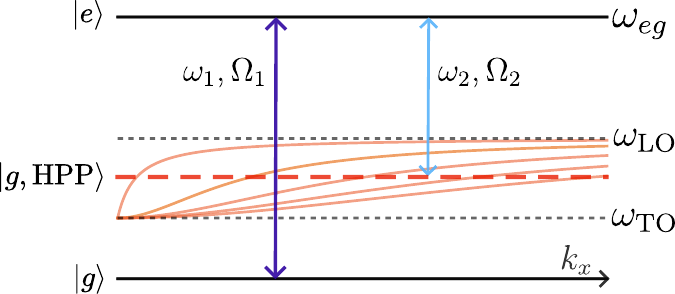}
\caption{\label{fig:ray_illus} A two-laser Raman scheme drives the emitter from $\ket{g}$ to $\ket{e}$ with a resonant pump \(\omega_{1}\simeq\omega_{eg}\) (purple), while a second (Raman) laser at \(\omega_{2}\) (blue) stimulates emission into the HPP manifold at 
\(\omega_{\mathrm{HPP}}=\omega_{1}-\omega_{2}\). 
The orange curves show the HPP dispersion $\omega_{\boldsymbol{k}_x,n}$ bounded by TO and LO phonon frequency; the red dashed line marks the fixed emission frequency \(\omega_{\mathrm{HPP}}\) set by \(\omega_{2}\). 
All branches satisfying $\omega_{\boldsymbol{k}_x,n}=\omega_{\mathrm{HPP}}$ are launched simultaneously, so tuning \(\omega_{2}\) selects the emission frequency, and hence the modal composition and propagation properties of the generated HPPs.}
\end{figure}

We can perform a rotating-wave transformation with $\omega_{eg}$ and the SW transformation as in Eq.~\eqref{eq:SW1}, the Hamiltonian becomes
\begin{eqnarray}
\hat{H}_{\rm sti}/\hbar&&\approx  \sum_{\boldsymbol{k_x},n}\omega_{\boldsymbol{k_x},n}\hat{\gamma}^\dagger_{\boldsymbol{k_x},n}\hat{\gamma}_{\boldsymbol{k_x},n}- \sum_{\boldsymbol{k_x},n} \frac{(\mu_d  E^{\rm vac}_x)^2}{\hbar\omega_{\boldsymbol{k_x},n}}\hat{\sigma}_{ee}
\nonumber\\
&&+[(\Omega_1e^{i\delta_1t}+\Omega_2e^{i\delta_2t})\hat{\sigma}_{eg}  + h.c.]\nonumber\\
&&-\sum_{\boldsymbol{k_x},n} \frac{\mu_d E^{\rm vac}_x }{\hbar\omega_{\boldsymbol{k_x},n}}(\hat{\gamma}_{\boldsymbol{k_x},n}-\hat{\gamma}^{\dagger}_{\boldsymbol{k_x},n})\nonumber\\
&&\times[(\Omega_1e^{i\delta_1t}+\Omega_2e^{i\delta_2t})\hat{\sigma}_{eg}  - h.c.],
\label{eq:H_sti_SW}
\end{eqnarray}
where $\delta_1=\omega_{eg}-\omega_{1}$, $\delta_2=\omega_{eg}-\omega_{2}$.
Assuming the first laser is at resonance $\delta_1\approx0$ and the second laser is detuned by a frequency in the hyperbolic regime $|\delta_2-\omega_{\boldsymbol{k_x},n}|\approx0\ll\delta_2, \omega_{\boldsymbol{k_x},n}$, we can do a second rotating-wave transformation with the HPP frequency $\omega_{\boldsymbol{k_x},n}$ and drop all fast-oscillating terms.
Then, last two terms in $H_{\rm sti}$ are reduced to
\begin{eqnarray}
\hat{H}_{\rm drive}&\approx&(\Omega_1\hat{\sigma}_{eg}+h.c.)\nonumber\\&-&\sum_{\boldsymbol{k_x},n} \frac{\mu_d E^{\rm vac}_x }{\hbar\omega_{\boldsymbol{k_x},n}} (\Omega_2 \hat{\sigma}_{eg}\hat{\gamma}_{\boldsymbol{k_x},n} - h.c.),
\label{eq:Raman_strength}
\end{eqnarray}
which is similar to the resonant Raman process of a $\Lambda$-system.
As illustrated in Fig.~\ref{fig:ray_illus}, a first laser $\omega_1 = \omega_{eg}$ excites the emitter from the ground state $\ket{g}$ to the excited state $\ket{e}$, while a second laser $\omega_2$ stimulates the decay, returning the emitter to the ground state, while emitting a HPP simultaneously.
As a result, the excited state effectively mediates the transition from the emitter ground state $\ket{g}$ to the polariton state $\ket{g,\rm{HPP}}$.

In Eq.~\eqref{eq:Raman_strength}, the coupling between $\ket{g}$ and $\ket{e}$ is set by the pump strength $\Omega_1$, whereas the transition between $\ket{e}$ and the combined state $\ket{g,\rm HPP}$ is given by the Raman laser strength times the vacuum electric field over HPP frequency $\Omega_2\frac{\mu_d E^{\rm vac}_x }{\hbar\omega_{\boldsymbol{k_x},n}}$.
As laser amplitudes are readily tunable, $\Omega_2$ provides a direct handle to boost the conversion into polaritons well beyond the spontaneous PSB rate.
When the scattering event is complete, the emitter returns to the ground state without any information about the process, so we would consider this stimulated HPP generation as a parametric process.

The spatial profile of the emitted polaritons also distinguishes the stimulated scheme from spontaneous PSB emission.
One unique feature of HPPs is that their electric field is highly focused, in a ray-like shape, which has been observed in the previous SNOM-type experiments \cite{daiSubdiffractionalFocusingGuiding2015a, liHyperbolicPhononpolaritonsBoron2015,leeImagePolaritonsBoron2020,herzigsheinfuxHighqualityNanocavitiesMultimodal2024}.
From Eq.~\eqref{eq:kxkz_ratio}, the direction of the wavevector of HPPs is governed by the ratio of the components of the dielectric function.
Since HPPs are longitudinal in the subwavelength limit, their electric field is spatially focused along the direction of the wavevector, forming a “ray” of polaritons.
In our point-source geometry, HPPs can be launched along any in-plane direction, so that the commonly used term ``ray-like'' corresponds more precisely to a cone of propagation in real space.
However, this ray-like behavior is likely obscured in spontaneous emission, where HPP frequency span the full Reststrahlen band.
The dielectric function ratio varies with frequency and therefore the propagation angles average out, washing away directional features. 
In contrast, stimulated emission locks the polariton frequency to a narrow linewidth, so the launched modes share a common angle and maintain phase over many reflections, producing a clear ray that propagates a longer distance. 
We analyze HPP propagation in more detail in Sec.~\ref{sec:propagation}.

In summary, this stimulated Raman transition offers several additional control knobs for the HPP generation, tuning the Raman frequency selects $\omega_{\rm HPP}$ within the Reststrahlen band and therefore sets the propagation angle of the HPP ray. 
Varying the Raman laser power controls the generation rate through the effective coupling. 
Adjusting the pulse duration (equivalently, bandwidth) switches between two regimes: laser pulses shorter than the emitter lifetime can produce single-HPP events analogous to spontaneous emission but at an enhanced rate, whereas narrowband pulses with duration longer than the emitter lifetime generate a coherent HPP ray at a fixed frequency.

\section{Propagation of Hyperbolic Phonon Polaritons}
\label{sec:propagation}

Having established two mechanisms for HPP generation, we now turn to the question of how these polaritons propagate once launched.
Understanding the spatial structure and coherence of HPPs is crucial for evaluating their suitability in quantum photonic applications, such as information transfer between emitters. 
In particular, we focus on the directional, ray-like propagation enabled by stimulated emission and investigate how this behavior depends on different parameters.

\begin{figure}
  \centering

  \subfloat[]{
    \includegraphics[width=0.8\columnwidth]{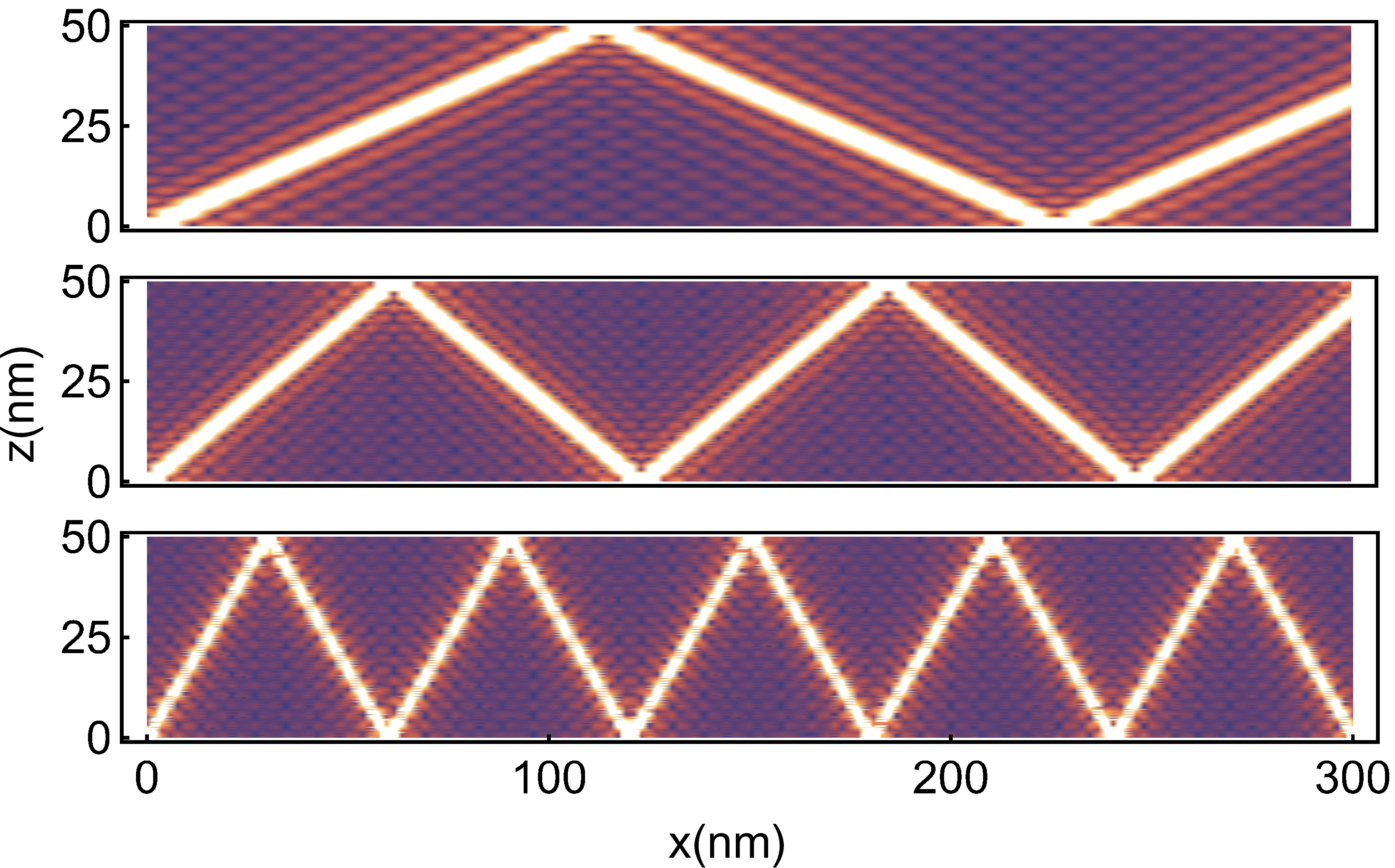}
    \label{fig:ray_freq}
  }\vspace{-0.5cm}
  \subfloat[]{
    \includegraphics[width=0.8\columnwidth]{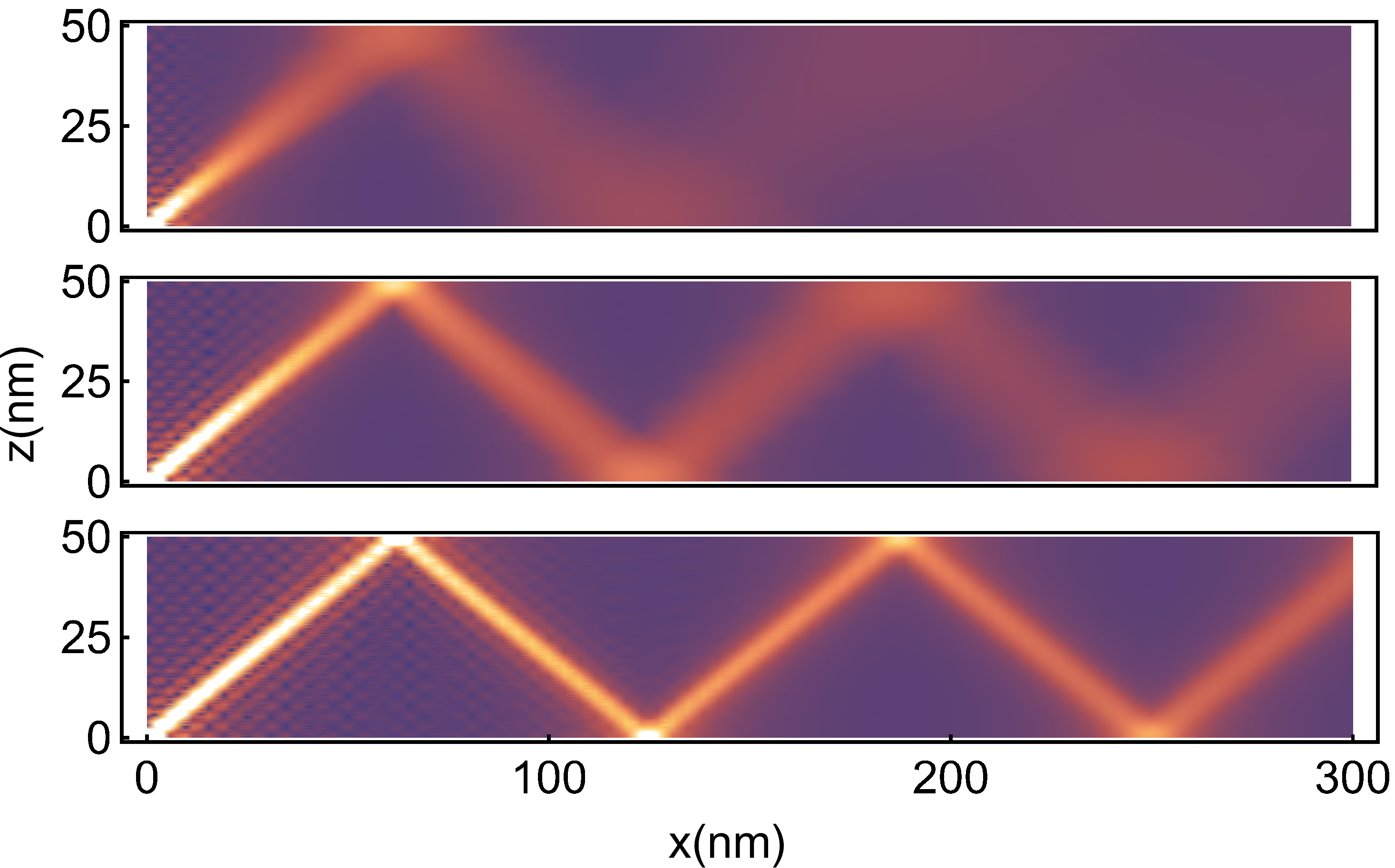}
    \label{fig:ray_delta}
  }
  \caption{\label{fig:ray_combined}
  Illustrations of the spatial distribution of the electric-field strength of HPPs, emitted from $\boldsymbol{r}=0$.
  Parameters used are $N_0=20$ and $d=50$ nm for illustration. 
  (a) At single frequencies: $\omega_{\mathrm{HPP}} =$ 43, 45, and 47 THz, from top to bottom. 
  The ray’s propagation angle depends on frequency through $\kappa(\omega)$.
  (b) With finite frequency width: $\delta=$ 500, 200, and 50 GHz, from top to bottom, centered at 45 THz.
  The sharpness of the ray diminishes more rapidly with increasing $\delta$.}
\end{figure}

As discussed above, stimulated emission from a color center can excite multiple HPP branches that share the same frequency.
We assume that these branches are excited uniformly up to a cutoff $N_0$, which limits the highest accessible branch index $n$.
Physically, $N_0$ corresponds to a momentum cutoff and can be estimated from Eq.~\eqref{eq:HPP_resonance} as
$N_0 \sim \frac{\Lambda d}{\kappa\pi}$.
The total electric field in real space is therefore given by a superposition of the mode functions of all excited HPP branches.
As shown by Eq.~\eqref{eq:monochromatic} in Appendix~\ref{app:ray}, constructive interference occurs when
\begin{eqnarray}
\kappa(\omega)x \pm z = 2dm,
\label{eq:ray_condition}
\end{eqnarray}
where $(x,z)$ are real-space coordinates and $m$ is an arbitrary integer.
This superposition produces spatially localized field maxima that trace out straight-line trajectories, referred to as HPP rays \cite{daiSubdiffractionalFocusingGuiding2015a,liHyperbolicPhononpolaritonsBoron2015,leeImagePolaritonsBoron2020,herzigsheinfuxHighqualityNanocavitiesMultimodal2024}.
These trajectories indicate that HPPs propagate at fixed angles determined by the frequency through $\kappa(\omega)$.
When HPPs are generated by a localized emitter, the width $w_{\mathrm{ray}}$ of a single-frequency HPP ray is set by the emitter size (Eq.~\eqref{eq:initial_width}).
Fig.~\ref{fig:ray_freq} shows representative electric-field distributions at different HPP frequencies.
For illustration, we take $N_0=20$ and $d=50~\mathrm{nm}$.
And we include a finite damping rate $\gamma_i$ in the dielectric function Eq.~\ref{eq:dielectric_function} to account for material losses in hBN ~\cite{gilesUltralowlossPolaritonsIsotopically2018}.


We now investigate how the spectral linewidth $\delta$ of the emitted polaritons influences the formation and propagation of the ray-like electric field pattern.
Physically, $\delta$ is determined by the characteristics of the external driving field, such as the pulse duration, or by intrinsic properties including the lifetime of the color center.
Fig.~\ref{fig:ray_delta} shows the electric field distribution for HPPs with linewidth $\delta=$ 500 GHz, 200 GHz, and 50 GHz. 
As expected, a broader linewidth leads to faster degradation of the ray, with the field pattern disappearing over shorter propagation distances.
This further supports the notion that sharp, directional rays cannot be formed via spontaneous decay, where the emission inherently spans the entire Reststrahlen band. We emphasize that the color centers could have radiatively broadened linewidths of a few GHz: consequently, it is possible to generate a ray of a single HPP pulse, with $100$~ps long Raman laser pulses, that would propagate over a micrometer length scale without significant broadening.

It is important to note that in Fig.~\ref{fig:ray_combined}, we assume a confinement along the $y$-direction for illustrative purposes.
In practice, however, the HPPs generated by a single color center propagate in the 2D $x$--$y$ plane, forming a polariton "cone" rather than a single ray.
This introduces geometric spreading with distance from the source (in the lossless limit one expects the field amplitude to scale as $|E|\propto 1/\sqrt{x}$).
In our lossy slab and for finite spectral linewidth, the decay is further enhanced by absorption and ray broadening.

\begin{figure}
\centering
\includegraphics[width=0.9\linewidth]{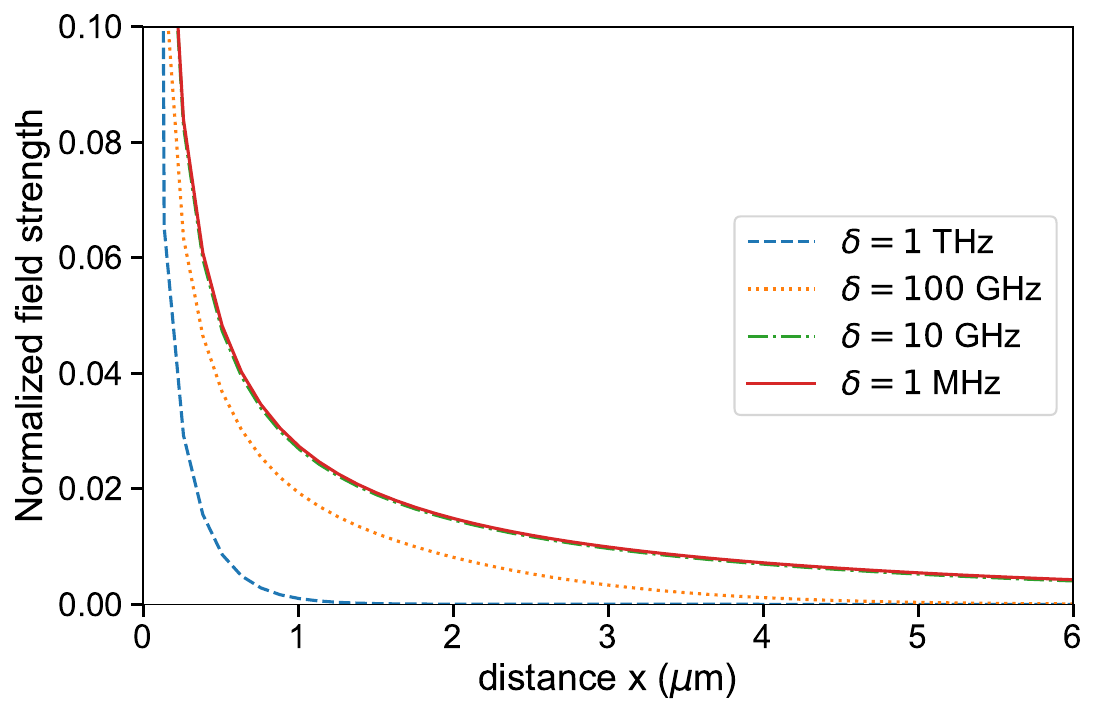}
\caption{\label{fig:ray_strength} Electric field amplitude of HPPs as a function of propagation distance for different linewidths: $\delta=$1 THz, 100 GHz, 10 GHz, 1 MHz. 
Field amplitudes are normalized to the value at $x = 10\,\text{nm}$.
For a broader linewidth, the field decays in a shorter distance.
Narrow linewidths ($\delta<10$ GHz) enable long-range propagation over several micrometers.
The distance dependence reflects the combined effects of geometric spreading, material loss, and linewidth-induced broadening.}
\end{figure}

To quantify this behavior, we calculate the electric field amplitude of HPPs as a function of propagation distance for various linewidths, as shown in Fig.~\ref{fig:ray_strength}.
For broad linewidths such as $\delta=$1 THz, the field diminishes within 1 $\mu m$. 
In contrast, for narrow linewidths below 10 GHz, the polariton field can persist over several micrometers with a power-law decay.
This highlights the importance of narrowband excitation for achieving long-range, directional polariton propagation.

In principle, HPPs should propagate over longer distances if we can impose in-plane confinement. 
One route is to use bound-state-in-the-continuum designs that localize HPPs in the plane, as demonstrated in Ref.~\cite{herzigsheinfuxHighqualityNanocavitiesMultimodal2024}. 
Such confinement converts the 2D polariton cone into guided modes and suppresses geometric spreading.
This would extend the propagation length, enhance emitter–HPP coupling, and effectively realize a waveguide for HPPs. 
Beyond simple confinement, another interesting possibility is to create an HPP resonator by tailoring the in-plane geometry. 
For example, Ref.~\cite{narimanovHyperbolicQuantumProcessor2024} proposed an elliptical hBN resonator that focuses HPP rays to mediate long-range interactions between spatially separated impurities, suitable for implementing quantum gates.
These structures suggest that HPPs in hBN could serve as building blocks for solid-state quantum devices, enabling controlled transport of energy and, in the quantum regime, the transfer of quantum states and information.

An experiment design to detect these HPP rays is similar to what we show in Fig.~\ref{fig:two_center}, to probe their ability to mediate interactions between two spatially separated emitters: one acting as the HPP source and the other as a detector \cite{cortesSuperCoulombicAtomAtom2017,newmanObservationLongrangeDipoledipole2018}.
The in-plane positions (x) of the hBN color centers can be measured with nanometer scale accuracy using an avalanche photodiode. However,  the out-of-plane (z) coordinates  are typically unknown.
We assume that the first emitter is driven by two lasers with frequencies $\omega_1 = \omega_{eg}$ and $\omega_2$, generating the HPP rays at frequency $\omega_{\rm HPP} = \omega_1 - \omega_2$.
The second emitter, several micrometers away, is illuminated only by the second laser at $\omega_2$. 
Only if the HPP ray intersects the second emitter, i.e. the distance between the two emitters satisfies Eq.~\eqref{eq:ray_condition}, the local field at $\omega_{\rm HPP}$ can resonantly drive it to the excited state, resulting in enhanced photoluminescence at $\omega_{eg}$.
By scanning the laser frequency $\omega_2$ and monitoring the PL intensity of the second emitter, one can confirm the presence of HPP rays and even extract information about the relative vertical (z) positions of both emitters.

\section{Conclusion}
\label{sec:conclusion}

In this study, we developed a cavity-QED framework to describe the interaction between color centers and hyperbolic phonon polaritons (HPPs) in hexagonal boron nitride (hBN), demonstrating how such emitters can serve as compact quantum sources of HPPs.
We analyzed two ways to generate HPPs: spontaneous emission via the phonon sideband (PSB) and stimulated Raman scattering.
The spontaneous emission can yield single-HPP events, and as the slab thickness decreases, the coupling concentrates into the lowest branch, producing a Purcell-like increase in the ultrathin limit.
We also proposed an experimental two-emitter setup that tests the single-polariton character through  photoluminescence correlation measurements.
In the stimulated case, the Raman process provides additional control over HPP generation-selecting the frequency, tuning the transition rate, and balancing quantum statistics with long-range coherence.
The analysis further shows that the polariton linewidth set by the drive and lifetime is a key parameter for spatial behavior: sufficiently narrowband excitation yields directional, ray-like propagation over micrometer distances.

Taken together, these results establish color centers as a powerful, intrinsically quantum complement to near-field techniques for creating and controlling HPPs.
Their atomic-scale localization compensates the momentum mismatch, enabling the transfer of their quantum properties to the photon field.
From the polaritonic perspective, integrating quantum emitters further strengthens HPPs in hBN as a practical platform for strong, and potentially ultrastrong, light–matter coupling in the mid-infrared.
On the color-center side, the long-range, ray-like HPPs provide a means for spatially separated emitters to interact.
In particular, the same concepts can be extended to spin-active defects \cite{sternRoomtemperatureOpticallyDetected2022}, where HPP-mediated interactions could be used to generate and manipulate entanglement between remote spins.
This perspective broadens the opportunities within the rapidly developing hBN color-center landscape and points to concrete applications, including quantum state transfer, entanglement distribution, and logical operations between solid-state qubits.

\section{ACKNOWLEDGMENTS}
We thank Igor Khanonkin, Le Liu, Evgenii Narimanov, Tao Shi, Simone De Liberato, Zhiyuan Sun, Dante Kennes, and Angel Rubio for fruitful discussions.
The work of J.E. and A.I. was supported by the Swiss National Science Foundation (SNSF) under Grant No. $200020\_207520$. 
J.K.~gratefully acknowledges support from Dr.~Max R\"{o}ssler, the Walter Haefner Foundation, and the ETH Z\"{u}rich Foundation.
S.C. and E.D. were supported by the SNSF project $200021\_212899$ and by the Swiss State Secretariat for Education, Research and Innovation (SERI) under contract number UeM019-1.
E.D., J.E., and A.I. were also supported by SNSF Sinergia grant CRSII--222792.

\appendix

\section{Derivation of the HPP mode function}
\label{app:eigenmodes}

Given that hBN is a 2D material, most experiments on HPPs employ a thin hBN slab, as depicted in Fig.~\ref{fig:setup}, in which the field oscillates inside the slab and decays evanescently outside.
The slab thickness is $d$.
We assume that the dielectric constant above the hBN slab is $\epsilon_1$, and that of the substrate is $\epsilon_2$.
For an in-plane wavevector chosen along the $x$ axis, we write the eigenmode profiles in terms of mode functions $F_x$ and $F_z$.

\begin{subequations}
\label{eq:modefunctions}
\begin{align}
&E_x= \sqrt{\frac{\hbar\omega}{2\epsilon_0}} F_x(x,z) \propto e^{i k_x x}
\left\{
\begin{array}{l}
    t_2 e^{-q_z (z-d)},\ (z>d)\\
    e^{ik_z z}+ r e^{-ik_z z},\\ (0\leq z \leq d)\\
    t_1 e^{q_z z},\ (z<0)
\label{eq:psix}
\end{array}
\right.
\\[-2pt]
&E_z = \sqrt{\frac{\hbar\omega}{2\epsilon_0}} F_z(x,z)  \propto  e^{i k_x x}
\left\{
\begin{array}{l}
    it_2 e^{-q_z (z-d)},\ (z>d)\\
    \frac{1}{\kappa}(e^{ik_z z}- r e^{-ik_z z}),\\ (0\leq z \leq d)\\
    -it_1 e^{q_z z}.\ (z<0)
\label{eq:psiz}
\end{array}
\right.
\end{align}
\end{subequations}
Here $t_1$, $t_2$, and $r$ are coefficients determined by electromagnetic boundary conditions at the two interfaces.
However, the overall magnitude of the mode function, i.e.\ $|F_x|$ (equivalently $|F_z|$), corresponds to the vacuum field strength and is therefore not fixed by the classical boundary-value problem.
It is determined by the quantization/normalization procedure.

In this work we consider slab thicknesses $d$ ranging from $1~\mathrm{nm}$ to $100~\mathrm{nm}$, which is far below the free-space wavelength (e.g.\ $40~\mathrm{THz}$ corresponds to $\sim 7.5~\mu\mathrm{m}$).
Accordingly, the characteristic HPP momentum inside hBN, of order $1/d$, is much larger than $\omega/c$, which justifies the quasi-static approximation.
Moreover, since the in-plane wavevector $k_x$ is conserved across the interfaces, the out-of-plane decay rate $q_z$ of the evanescent field outside the slab satisfies the (isotropic) dispersion relation
\begin{eqnarray}
k_x^2-q_z^2 = \epsilon_\mathrm{out}\frac{\omega^2}{c^2} \approx 0,
\label{eq:qz}
\end{eqnarray}
where $\epsilon_{out}$ denotes the permittivity of the outer medium (either $\epsilon_1$ or $\epsilon_2$).

Because there are two interfaces at $z=0$ and $z=d$, there are four independent boundary conditions.
Choosing the continuity of $E_x$ and $D_z$ at both interfaces provides four equations that determine the three coefficients $t_1$, $t_2$, and $r$, and yields an additional resonance condition that quantizes the wavevector along the $z$ direction:
\begin{eqnarray}
e^{2ik_z d}= \frac{(i\epsilon_1 - \kappa\epsilon_x)(i\epsilon_2 - \kappa\epsilon_x)}{(i\epsilon_1 + \kappa\epsilon_x)(i\epsilon_2 + \kappa\epsilon_x)}.
\label{eq:resonance1}
\end{eqnarray}
This condition can be rewritten in the more familiar form given in Eq.~\eqref{eq:HPP_resonance},
\begin{eqnarray}
k_x = \frac{\kappa}{d}[\text{tan}^{-1}(-\frac{\epsilon_1}{\epsilon_x\kappa}) + \text{tan}^{-1}(-\frac{\epsilon_2}{\epsilon_x\kappa}) + n\pi ],
\label{eq:resonanc2e}
\end{eqnarray}
where $n$ is the branch index.
The corresponding coefficients are
\begin{eqnarray}
&r = \frac{i\epsilon_x\kappa+\epsilon_1}{i\epsilon_x\kappa-\epsilon_1}, \
t_1 = 1+r, \
t_2 = e^{ik_zd}+re^{-ik_zd}.
\label{eq:rt1t2}
\end{eqnarray}
Finally, since $\epsilon_x$ and $\epsilon_z$ are frequency dependent, the resonance condition determines the dispersion relation of HPPs in the thin slab.

\section{Quantization of EM field in Material}
\label{app:quantization}

This is a longstanding problem in QED.
The most common approach is to calculate the classical EM modes for a given structure with Maxwell's equations and then quantizing these modes by normalizing their stored energy to that of a single photon (or vacuum photon) at the mode frequency.
While this method is widely used due to its simplicity, it has some limitations.

One issue is the difficulty to interpret the macroscopic Maxwell's solution in the polariton picture, where the eigenmode of the system should include both photon and matter degrees of freedom.
For example, for bulk phonon polaritons, this connection is intuitive: the classical EM modes represent the hybridization between TO phonons and photons.
However, in the case of HPPs, this intuitive explanation breaks down.
In particular, in the subwavelength (quasi-static) regime, the EM field is predominantly longitudinal, and the vector potential becomes negligible in the Coulomb gauge.

Another obstacle is dissipation.
While our initial efforts neglect material losses for simplicity, incorporating dissipation significantly complicates the problem.
In lossy materials, defining the energy density becomes nontrivial, adding further challenges to the quantization process \cite{feistMacroscopicQEDQuantum2020a}.

To address this, theoretical frameworks such as Macroscopic QED have been developed.
These approaches quantize the electromagnetic field in matter using the macroscopic dielectric function, which implicitly accounts for the material’s response, including dissipation and dispersion \cite{glauberQuantumOpticsDielectric1991,scheelMacroscopicQEDConcepts2009a,riveraLightMatterInteractions2020a}.

In our work, we present two methods for quantizing HPPs using the dielectric function $\epsilon_i(\omega)$ and show their equivalence.
Additionally, by neglecting losses, an approximation valid for low-loss hBN, we construct a microscopic Hamiltonian for phonon polaritons to confirm the results of Macroscopic QED and gain deeper insight into the nature of HPPs.

\subsection{Quantization with dispersive energy density}
A straightforward method to generalize field quantization in materials is to normalize the electromagnetic energy density $U(\omega)$ using its expression in dispersive media \cite{lifsicElectrodynamicsContinuousMedia1993}:
\begin{eqnarray}
U(\omega) = \frac{1}{2} [\frac{d(\omega\epsilon)}{d\omega}E^2+ \frac{d(\omega\mu)}{d\omega}H^2] ,
\label{eq:energydensity_landau}
\end{eqnarray}
where $E$ and $H$ are the electric and magnetic fields, respectively, and $\epsilon(\omega)$ and $\mu(\omega)$ are the frequency-dependent permittivity and permeability.
Assuming no magnetic response ($\mu = \mu_0$) and working in the Weyl (temporal) gauge, where the scalar potential vanishes and $\mathbf{E} = -\partial_t \mathbf{A}$, we can express the energy density in terms of the vector potential $\mathbf{A}$:
\begin{eqnarray}
U(\omega) = \frac{1}{2} \left (\frac{d(\omega\epsilon)}{d\omega}\omega^2 A^2+ \frac{(\nabla\times\boldsymbol{A})^2}{\mu_0}\right).
\label{eq:energydensity_A}
\end{eqnarray}
Then, we can quantize the vector potential in the standard way:
\begin{eqnarray}
\boldsymbol{\hat{A}}(\boldsymbol{r})=\sum_m \sqrt{\frac{\hbar}{2\epsilon_0\omega_m}}[\boldsymbol{F}_m(\boldsymbol{r})\boldsymbol{\hat{\gamma}}_m +\boldsymbol{F}^*_m(\boldsymbol{r})\boldsymbol{\hat{\gamma}}^\dagger_m],
\label{eq:quantized_A}
\end{eqnarray}
where $\boldsymbol{F}_m$ denotes the m-th mode function of vector potential with mode frequency $\omega_m$, and $\boldsymbol{\hat{\gamma}}^\dagger_m(\boldsymbol{\hat{\gamma}}_m)$ is the corresponding polariton creation (annihilation) operator.
The mode functions $\mathbf{F}_m(\mathbf{r})$ are obtained from solutions of Maxwell’s equations and the normalization condition in subwavelength limit is \cite{archambaultQuantumTheorySpontaneous2010,riveraLightMatterInteractions2020a}
\begin{eqnarray}
\frac{1}{2\omega_m}\int d^3\boldsymbol{r} \boldsymbol{F}^*_m(\boldsymbol{r})\frac{d(\omega^2\boldsymbol{\epsilon})}{d\omega}|_{\omega=\omega_m}\boldsymbol{F}_m(\boldsymbol{r}) = 1,
\label{eq:normalization_energydensity}
\end{eqnarray}
where $\boldsymbol{\epsilon}(\omega)$ is the dielectric tensor of the material.
This formalism holds for dispersive but lossless media, such as high-purity hBN.
Extensions to include lossy media have also been discussed in the literature \cite{knollQEDDispersingAbsorbing2003,scheelMacroscopicQEDConcepts2009a}.
For HPPs in a thin slab, $m$ should be replaced by the in-plane wavevector $k_x$ and out-of-plane quantum number $n$, and $\omega_{k_x,n}$ is shown in Fig.~\ref{fig:dispersion_HPP}.

\subsection{Quantization with Green function}
Another interesting framework is that normalizing the mode function by matching the quantum Green function with classical Green function from Poisson equation.
We refer to the work of Ref.\cite{andolinaQuantumElectrodynamicsGraphene2025a} to quantize HPPs, where the electric field is described by a scalar potential.
As mentioned, this is based on the well-known quasi-static approximation of subwavelength electromagnetic field that the electric field becomes predominantly longitudinal and can be expressed as the gradient of a quasi-static potential $\phi$, while the magnetic field is negligible \cite{bergmanSurfacePlasmonAmplification2003,feistMacroscopicQEDQuantum2020a}: 
\begin{eqnarray}
\boldsymbol{E} = -\nabla\phi.
\label{eq:quasistatic}
\end{eqnarray}
We can quantize $\phi$ in the following form,
\begin{eqnarray}
\phi(\boldsymbol{r})= \sum_m \sqrt{\frac{\hbar\omega_m}{2\epsilon_0}}[f_m(\boldsymbol{r})\boldsymbol{\hat{\gamma}}_m +f_m^*(\boldsymbol{r})\boldsymbol{\hat{\gamma}}_m^\dagger],
\label{eq:quantized_phi}
\end{eqnarray}
where $f_m$ is the mode function for scalar potential $\phi$.
The corresponding electric field mode function is given by $\boldsymbol{F}_m=-\nabla f_m$.
The normalization condition derived by matching the quantum and classical Green functions is \cite{andolinaQuantumElectrodynamicsGraphene2025a}
\begin{eqnarray}
\frac{\omega_m}{2}\int d^3\boldsymbol{r} f_m^*(\boldsymbol{r}) (k_x^2\frac{d\epsilon_x}{d\omega}+ k_z^2\frac{d\epsilon_z}{d\omega})|_{\omega=\omega_m}f_m(\boldsymbol{r}) = 1. 
\nonumber\\
\label{eq:normalization_greenfucntion}
\end{eqnarray}
This condition is equivalent to Eq.~\eqref{eq:normalization_energydensity}, if $\int d^3\boldsymbol{r} f_m^*(\boldsymbol{r}) (k_x^2\epsilon_x+ k_z^2\epsilon_z) f_m(\boldsymbol{r}) = 0$, which always holds in the subwavelength limit due to the TM-mode dispersion relation Eq.~\eqref{eq:TM_dispersion}.

\subsection{Quantization with microscopic Hamiltonian}

The Lagrangian density describing the anisotropic phonon interacting with the electromagnetic field is given by \cite{gubbinRealspaceHopfieldDiagonalization2016,sentefCavityQuantumelectrodynamicalPolaritonically2018,ashidaQuantumElectrodynamicControl2020}
\begin{eqnarray}
\mathcal{L}_0 =&& \sum_{i\in\{x,z\}} \frac{1}{2}\epsilon_0\epsilon_{i\infty}(\dot{\boldsymbol{A}}+\nabla \phi )_i^2- \frac{1}{2\mu_0}(\nabla\times\boldsymbol{A})^2 \nonumber\\
&&+ \sum_{i\in\{x,z\}} \frac{1}{2}\frac{M_i}{v_{uc}} \dot{\boldsymbol{x}}^2-\sum_{i\in\{x,z\}}\frac{1}{2}\frac{M_i}{v_{uc}}\Omega_{TO,i}^2x_i^2 \nonumber\\
&&+\frac{e^*}{v_{uc}} \dot{\boldsymbol{x}}\cdot\boldsymbol{A} - \frac{e^*}{v_{uc}}\boldsymbol{x}\cdot(\nabla \phi),
\label{eq:general_L}
\end{eqnarray}
where $\boldsymbol{A}$ is the vector potential, $\phi$ is the scalar potential, and $\boldsymbol{x}$ is the phonon displacement.
$i$ represents the in-plane or out-of-plane components.
$\Omega_{TO}(\Omega_{LO})$ is the TO (LO) phonon frequency and $\epsilon_{\infty}$ is the high-frequency dielectric constant.
$\epsilon_0$ is the vacuum permittivity, $e^*$ is the effective charge, $M$ is the reduced mass of ions and $v_{uc}$ is the unit cell volume.
To proceed with quantization, we perform a Legendre transformation to derive the Hamiltonian. 
The conjugate momenta are defined as
\begin{eqnarray}
&&\Pi_i=\frac{\delta\mathcal{L}_0}{\delta\dot{A}_i}= \epsilon_0\epsilon_{i\infty} \dot{A}_i, \\
&&\pi_i=v_{uc} \frac{\delta\mathcal{L}_0}{\delta\dot{x}_i}= M_i\dot{x}_i+e^*A_i.
\label{eq:coul_conjugate}
\end{eqnarray}
To further simplify, we adopt the Weyl gauge, assuming the scalar potential is zero, $\phi = 0$.
The corresponding Hamiltonian is 
\begin{eqnarray}
H_{weyl}= \int d\boldsymbol{r} \sum_{i\in\{x,z\}}\frac{\Pi_i^2}{2\epsilon_0\epsilon_{i\infty}} + \frac{\epsilon_0 c^2}{2} (\nabla\times \boldsymbol{A})^2 \nonumber\\
+ \sum_{i\in\{x,z\}} \frac{(\boldsymbol{\pi} - e^* \boldsymbol{A})_i^2}{2 M_iv_{uc}} + \sum_{i\in\{x,z\}} \frac{1}{2} \frac{M_i}{v_{uc}} \Omega_{TO,i}^2 x_i^2.
\label{eq:weyl_H}
\end{eqnarray}
The eigenmodes of the system can be obtained by diagonalizing this Hamiltonian \cite{gubbinRealspaceHopfieldDiagonalization2016,ashidaCavityQuantumElectrodynamics2023}. 
The resulting normalization condition is
\begin{eqnarray}
\sum_{i\in\{x,z\}} \int d^3\boldsymbol{r}&&[ 1 + \frac{g_{ph,i}^2\Omega^2_{i,TO}}{(\Omega_{TO,i}^2 - \omega^2)^2}\Theta(z)\Theta(d-z)]\nonumber\\
&&\times F^*_{m,i}(\boldsymbol{r}) F_{m,i}(\boldsymbol{r}) = 1,
\label{eq:normalization_digonalization}
\end{eqnarray}
Here, the first term represents the contribution from the free electromagnetic field, while the second term arises from the electric field induced by phonons inside the hBN slab, where $\Theta$ is the Heaviside step function and $g_{ph,i}=\sqrt{\frac{(e^*)^2}{\epsilon_{i,\infty}M_i v_{uc}}}$.
Notably, the same normalization condition can be derived within the macroscopic QED framework in Eq.~\eqref{eq:normalization_energydensity} using the Lorentzian dielectric function, demonstrating consistency between the microscopic and macroscopic approaches.

\subsection{HPPs and LO phonons}

Fig.~\ref{fig:vacuum_strength} shows the in-plane vacuum fluctuation of the electric field in an hBN slab.
For conventional (bulk) phonon polaritons, the hybridized modes become increasingly phonon-like in the large-momentum limit, and the associated electric field typically decreases with larger wavevector.
In contrast, HPPs exhibit a saturation of the in-plane electric field at large $k_x$.
This behavior stems from the longitudinal character of HPPs in the subwavelength regime.
It is similar to surface phonon polaritons, where strong hybridization between photonic and phononic components persists even at large momenta.

In fact, in the limit $k_x d \to \infty$ ($\omega\to \Omega_{x,\rm LO}$), the electric field for a specific HPP branch (with finite $n$) is purely induced by the in-plane LO phonon of hBN,
\begin{eqnarray}
E^{\rm vac}_x \bigg|_{\substack{k_x d \to \infty \\ \omega \to \Omega_{x,\mathrm{LO}}}}
&\longrightarrow&
\frac{e^*}{\epsilon_0 \epsilon_\infty v_{\mathrm{uc}}} \left| \boldsymbol{x}_{\mathrm{LO}} \right|
\nonumber \\
&=&
\frac{e^*}{\epsilon_0 \epsilon_\infty v_{\mathrm{uc}}}
\sqrt{ \frac{\hbar}{2 M_x \Omega_{x,\mathrm{LO}} V_{\mathrm{eff}} / v_{\mathrm{uc}}} } .\nonumber\\
\label{eq:LO_HPPS}
\end{eqnarray}
Here $\left|\boldsymbol{x}_{\mathrm{LO}}\right|$ denotes the vacuum fluctuation of the LO-phonon displacement within an effective volume $V_{\mathrm{eff}}$, explicitly demonstrating the phononic nature of HPPs in this regime \cite{riveraPhononPolaritonicsTwoDimensional2019,michelPhononDispersionsPiezoelectricity2011}.

\section{Photoluminescence Excitation Setup}
\label{app:PL_setup}

\begin{figure}
\centering
\includegraphics[width=0.8\linewidth]{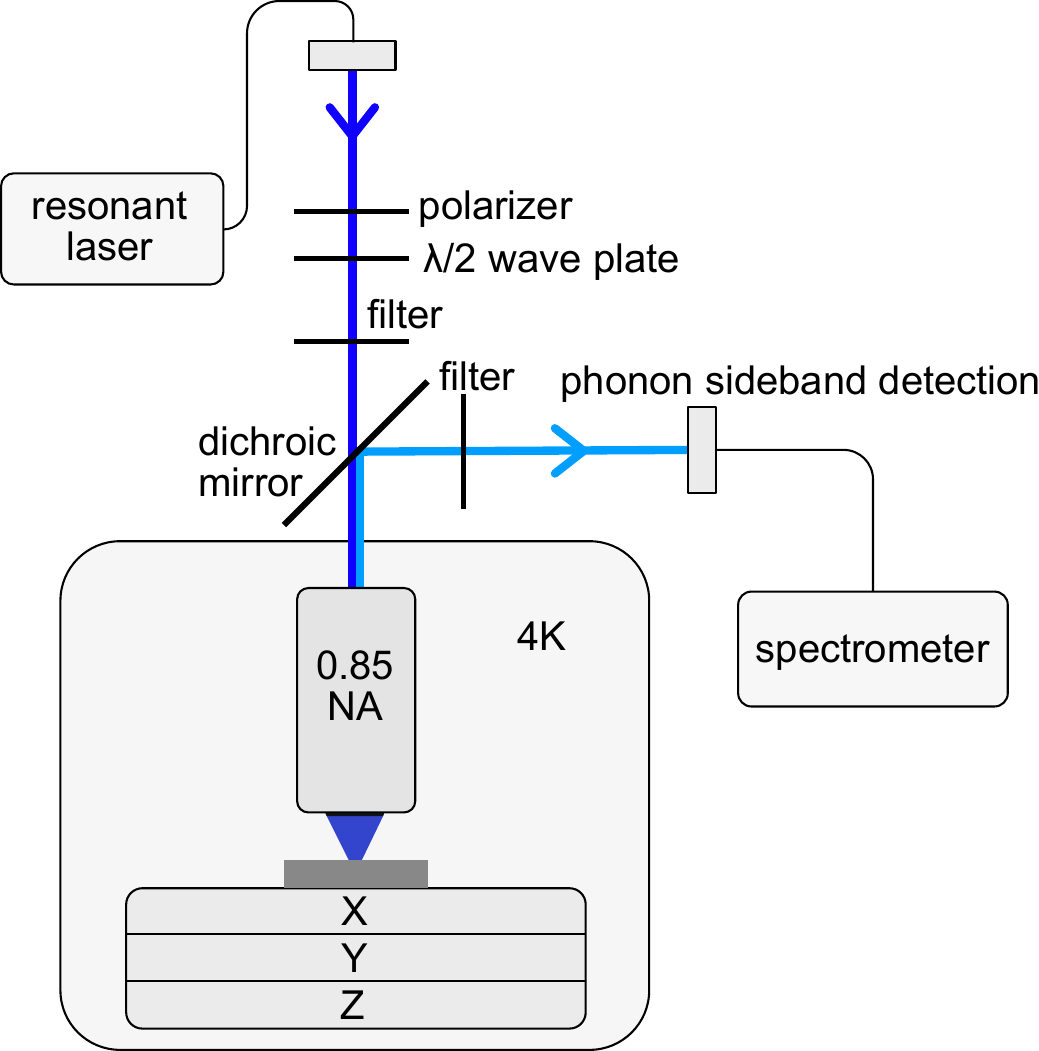}
\caption{\label{fig:PL_setup} 
The photoluminescence excitation setup uses a frequency-doubled Ti:Sapphire laser tuned to the B-center resonance. A polarizer and half-wave plate allow for polarization control, and the laser is spectrally cleaned by a narrow band pass filter. A dichroic mirror separates excitation and detection paths, and a 0.85 NA objective focuses the laser beam onto the B-centers while collecting the emitted light. The phonon side band is isolated by a band pass filter and measured using a spectrometer.}
\end{figure}

The photoluminescence excitation setup is shown in Fig.~\ref{fig:PL_setup}. A Ti:Sapphire laser (Sirah Matisse CR) is frequency-doubled using a PPLN crystal (Covesion) tuned to the B-center resonance. Emission is measured with a spectrometer (Princeton Instruments, Inc.) with a 1500~g\,mm$^{-1}$ optical grating. The sample is positioned using Attocube XYZ positioners and an XY scanner, and cooled to 4K in a liquid-helium bath cryostat.

\section{Additional Photoluminescence Calculations}
\label{app:additional_PL}
Here we present more results of the photoluminescence calculation.

\begin{figure}
\centering
\includegraphics[width=0.9\linewidth]{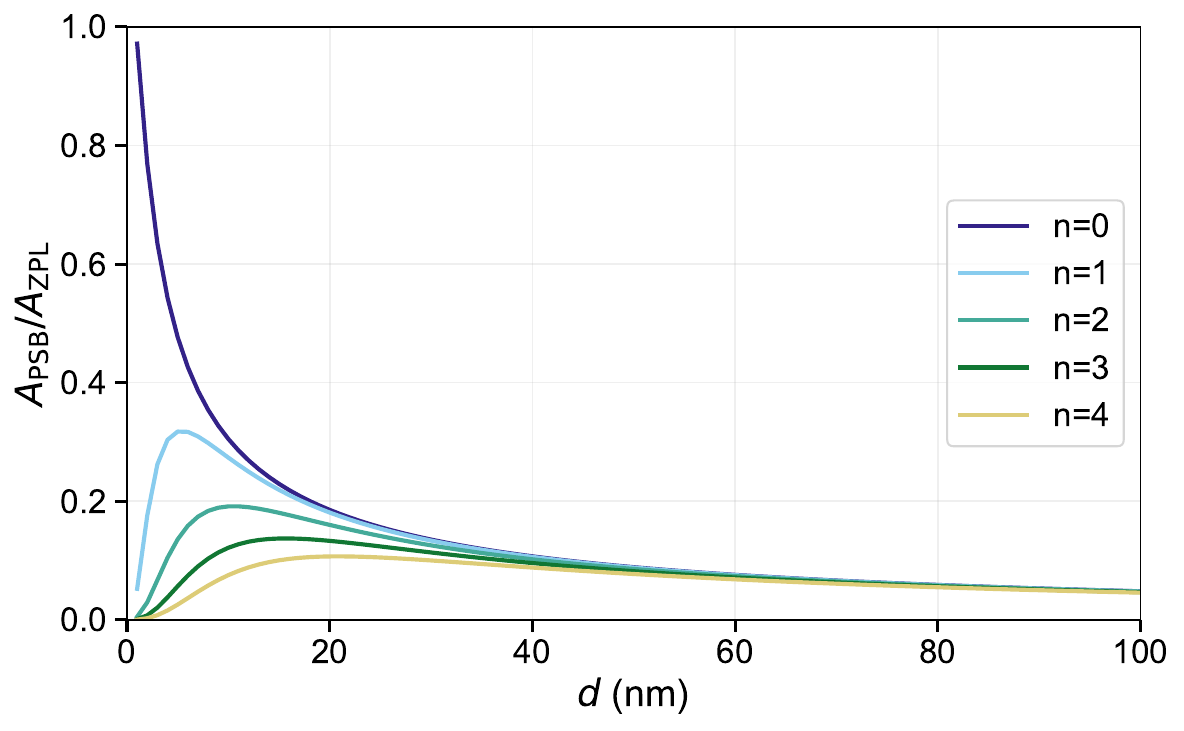}
\caption{\label{fig:PSB_Lambda1nm} 
Calculated intensity ratio between PSB and ZPL as a function of slab thickness. 
We use a momentum cutoff \( \Lambda = 1/(1\,\text{nm}) \) and dipole moment \( \mu_d = e_0 \cdot (2\,\text{nm}) \) for illustration. 
Solid curves correspond to HPP branches with mode indices \( n = 0, 1, 2, 3, 4 \) (from top to bottom).}
\end{figure}

Fig.~\ref{fig:PSB} shows the calculated PSB/ZPL intensity ratio using a dipole moment $\mu_d=e_0\cdot(2\ \text{nm})$ and momentum cutoff for HPPs $\Lambda=1/(2\ \text{nm})$.
Within our treatment, varying only the dipole moment scales the coupling linearly, and therefore the ratio obeys $A_{PSB}/A_{ZPL} \propto \mu_d^2$.
On the other hand, increasing the cutoff $\Lambda$ enlarges the available phase space in the momentum sum of Eq.~\eqref{eq:PSB_ZPL_ratio}, and therefore increases the PSB intensity.
The result for $\mu_d=e_0\cdot(2\ \text{nm})$ and $\Lambda=1/(1\ \text{nm})$ is shown in Fig.~\ref{fig:PSB_Lambda1nm}.
We can see that the PSB is stronger than in the $\Lambda=1/(2\ \text{nm})$ case, while the qualitative thickness dependence and the features for different branches remain unchanged.

\begin{figure}
\centering
\includegraphics[width=0.9\linewidth]{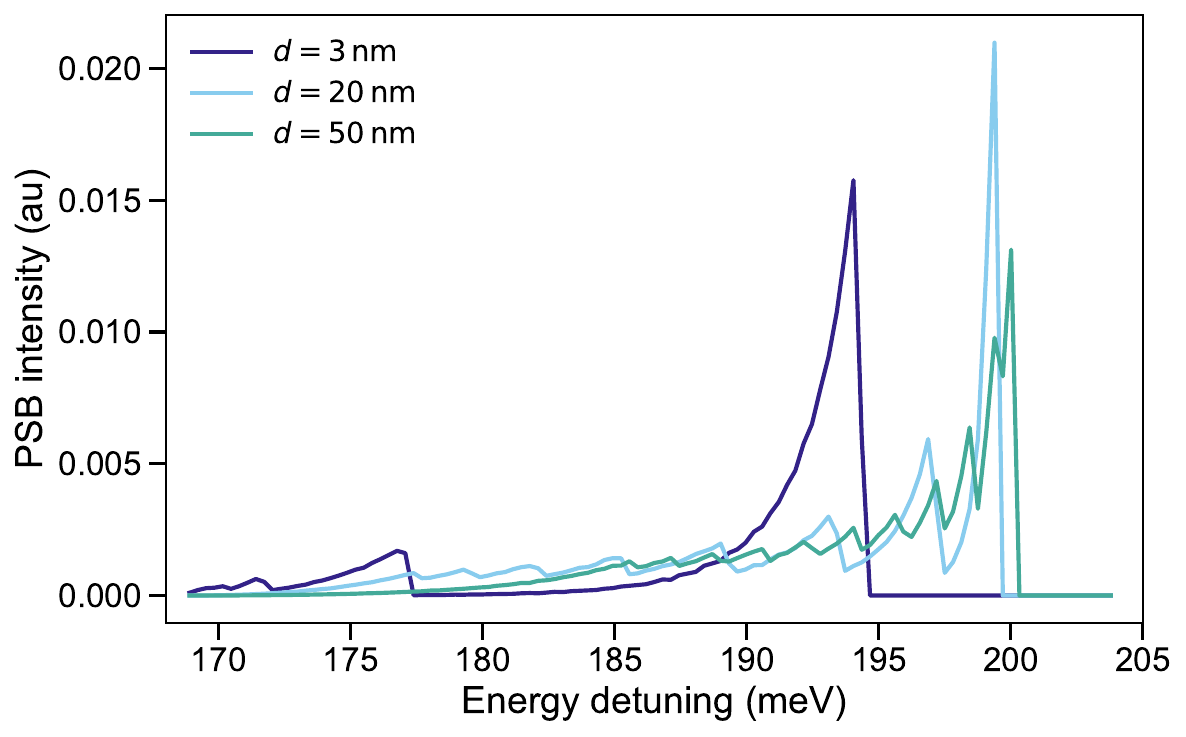}
\caption{\label{fig:PSB_spectrum_vary_d} 
Calculated PSB spectrum with different slab thickness. 
We use a momentum cutoff \( \Lambda = 1/(2\,\text{nm}) \) and dipole moment \( \mu_d = e_0 \cdot (2\,\text{nm}) \). 
The frequency resolution is set to be $\frac{\Omega_{x,LO}-\Omega_{x,TO}}{100}\approx76\ \text{GHz}\ (0.3\ \text{meV})$.
Solid curves correspond slab thickness $d=3, 20, 50\ \text{nm}$.}
\end{figure}

We also present the data of frequency-resolved PSB spectra in Fig.~\ref{fig:PSB_spectra}.
For comparison, Fig.~\ref{fig:PSB_spectrum_vary_d} presents calculated spectra from our simplified model for several slab thicknesses $d$.
The spectra exhibit a sequence of narrow maxima associated with individual HPP branches, appearing where the HPP density of states is largest.
As the slab becomes thinner, the peak of the lowest branch ($n=0$) stands out, and the PSB maximum shifts to lower energy.
This trend is consistent with the experimentally observed second PSB feature occurring below the in-plane LO energy.
In the opposite limit, the spectrum approaches the bulk response: the peak positions move toward the in-plane LO frequency and the HPP description reduces to the in-plane LO-phonon response (Eq.~\eqref{eq:LO_HPPS}).
From this viewpoint, assignments of the second feature to Fr\"ohlich coupling with LO phonon in previous studies are recovered in the bulk limit~\cite{wiggerPhononassistedEmissionAbsorption2019,khatriPhononSidebandsColor2019}.

\section{Derivation of the HPP ray}
\label{app:ray}
We know the mode function of HPPs from Appendix \ref{app:eigenmodes}.
If we can excite multiple HPP branches with a single frequency, we only need to sum over $n$ to obtain the field distribution inside the hBN slab. 
\begin{eqnarray}
&&E_x(x,z,\omega)\propto \Sigma_{n=0}^{N_0} (e^{i k_z z}+re^{-ik_z z})e^{ik_x x }
\nonumber \\
&&= \Sigma_{n=0}^{N_0} [e^{i(k_0+\frac{n\pi}{d})z}+re^{-i(k_0+\frac{n\pi}{d})z}]e^{i\kappa(k_0+\frac{n\pi}{d})x} \nonumber\\
&&= e^{ik_0 (\kappa x+z)}
\frac{1-e^{i\frac{N_0\pi}{d}(\kappa x+z)}}{1-e^{i{\frac{\pi}{d}(\kappa x +z)}}}\nonumber + re^{ik_0 (\kappa x-z)}
\frac{1-e^{i\frac{N_0\pi}{d}(\kappa x-z)}}{1-e^{i{\frac{\pi}{d}(\kappa x -z)}}},\\
&&k_0=\frac{2}{d}[\text{tan}^{-1}(-\frac{1}{\epsilon_x\kappa})],
\label{eq:monochromatic}
\end{eqnarray}
where cutoff $N_0$ sets the maximum value of $n$.
This monochromatic field exhibits a well-defined ray-like pattern, with the field strength peaking (and diverging in the lossless case) at the poles of the denominator, as suggested in Eq.~\eqref{eq:ray_condition}.

The transverse width $w_{\rm ray}$ of a single-frequency HPP ray is determined by how many modes interfere constructively within the ray.
Using $N_0 \simeq \Lambda d/\kappa\pi$, we obtain
\begin{align}
    w_{\rm ray} \approx \frac{d}{N_0 \kappa}
    \simeq \frac{\pi}{\Lambda}
    \sim l_e ,
\label{eq:initial_width}
\end{align}
so, the initial ray width is set by the emitter size $l_e$ through the momentum cutoff $\Lambda \sim 1/l_e$.

Next, we consider a finite frequency width $\delta$ around the central frequency $\omega_0$.
To include this, we can integrate over frequency when computing the EM field.
The oscillatory phase factor $e^{i\omega t}$ is neglected since we are working in the quasi-static limit, where $\omega\ll kc$.

\begin{eqnarray}
E(x,z)= \int^{+\infty}_{-\infty} d\omega E(x,z,\omega) f(\omega)
\label{eq:omega_int}
\end{eqnarray}
where $f$ is the frequency distribution, and we just take the gaussian distribution for simplicity. 
\begin{eqnarray}
f(\omega)= \frac{1}{\sqrt{2\pi\delta^2}}e^{-\frac{(\omega-\omega_0)^2}{2\delta^2}}
\label{eq:gaussian}
\end{eqnarray}
The resulting field strength from Eq.~\eqref{eq:omega_int} is shown in Fig[\ref{fig:ray_delta}].
With a finite width $\delta$, the sharp ray-like character of the HPP gradually fades over distance.
As expected, a smaller $\delta$ preserves the focused propagation over a longer range.

In the above calculation, we assume the ray can only propagate along x direction.
However, HPPs generated by a single color center propagate isotropically in the 2D x-y plane, forming a polariton “cone” rather than a single ray. 
We should use the cylindrical coordinates to describe the EM field.

For simplicity, we work in the longitudinal or sub-wavelength limit from the start. 
In this regime, the electric field can be approximated as the gradient of a quasi-static scalar potential $U$:
\begin{eqnarray}
\boldsymbol{E}=-\nabla U
\label{eq:quasi_scalar}
\end{eqnarray}
In cylindrical coordinates, Maxwell’s equations in this limit yield :
\begin{eqnarray}
[\frac{1}{\epsilon_z}(\frac{1}{\rho}\frac{\partial}{\partial\rho}+\frac{\partial^2}{\partial\rho^2})+\frac{1}{\epsilon_t}\frac{\partial^2}{\partial z^2}]U=0
\label{eq:dispersion}
\end{eqnarray}
where $\rho$ is the radial coordinate and $z$ is the out-of-plane axis coordinate.
$\epsilon_t(\epsilon_z)$ is the in-plane (out-of-plane) component of the dielectric function.
The general solution takes the form
\begin{eqnarray}
U \propto J_0(k_t\rho)(e^{ik_z z}+re^{-ik_z})
\label{eq:soln}
\end{eqnarray}
where $J_0$ is the Bessel function.
Under this form, the Laplacian equation becomes equivalent to that in Cartesian coordinates:
\begin{eqnarray}
\frac{k_t^2}{\epsilon_z}+\frac{k_z^2}{\epsilon_t}=0
\label{eq:dispersion2}
\end{eqnarray}
Since the boundary conditions still require continuity of both $U$ and $\epsilon_z\partial_zU$ across the interfaces, the dispersion relations and wavevector solutions remain unchanged.
\begin{eqnarray}
&&k_z(\omega,n)=k_0(\omega) + \frac{n\pi}{d}\nonumber\\ 
&&k_t(\omega,n)= \kappa(\omega)k_z(\omega,n)\nonumber\\
&&r(\omega) = \frac{i\epsilon_t(\omega)\kappa(\omega)+1}{i\epsilon_t(\omega)\kappa(\omega)-1}
\label{eq:wavevector}
\end{eqnarray}

However, to properly describe propagating HPPs, we should use the Hankel function $H^{(1)}_\alpha$ to describe quasi-static potential, instead of Bessel function of the first kind.
\begin{eqnarray}
U \propto H_0^{(1)}(k_t\rho)(e^{ik_z z}+re^{-ik_z})
\label{eq:ray_poten}
\end{eqnarray}
And for the electric field
\begin{eqnarray}
E_t \propto H_1^{(1)}(k_t\rho)(e^{ik_z z}+re^{-ik_z})
\label{eq:ray_ele}
\end{eqnarray}
This expression accurately describes the electric field of the HPPs except at the origin $\rho=0$, where the Hankel function diverges due to its non-analytic behavior at the origin in polar coordinates.

As before, we assume the phonon-assisted decay from the B-center will excite all modes $n$ of HPP evenly, to a cutoff $N_0$, at a single frequency.
\begin{eqnarray}
E(x,z,\omega) \propto \Sigma_{n=0}^{N_0} H_1^{(1)}(k_t\rho)(e^{ik_z z}+re^{-ik_z})
\label{eq:ray}
\end{eqnarray}
This is the equation used to calculate the result in Fig.~\ref{fig:ray_strength}.
The key difference compare to Eq.~\eqref{eq:monochromatic} is that, in cylindrical coordinates, the field strength decreases with radial distance, reflecting the spreading of the field over a larger area.

\bibliography{reference}

\end{document}